\documentclass[aps,showpacs]{revtex4}

\usepackage{graphicx}
\usepackage{dcolumn}
\usepackage{bm}

\begin{document}

\preprint{v88}

\title{\textbf{Relativity of GPS Measurement}}

\author{Thomas B. Bahder}
\email[]{bahder@arl.army.mil}
\affiliation{U. S. Army Research Laboratory \\
2800 Powder Mill Road \\
Adelphi, Maryland, USA  20783-1197}

\date{\today}

\begin{abstract}
The relativity of Global Positioning System (GPS) pseudorange measurements
is explored within the geometrical optics approximation in the curved
space-time near Earth. A space-time grid for navigation is created by the
discontinuities introduced in the electromagnetic field amplitude by the
P-code broadcast by the GPS satellites. We compute the world function of
space-time near Earth, and we use it to define a scalar phase function that
describes the space-time grid. We use this scalar phase function to define
the measured pseudorange, which turns out to be a two-point space-time
scalar under generalized coordinate transformations. Though the measured
pseudorange is an invariant, it depends on the world lines of the receiver
and satellite. While two colocated receivers measure two different
pseudoranges to the same satellite, they obtain correct position and time,
independent of their velocity. We relate the measured pseudorange to the
geometry of space-time and find corrections to the conventional model of
pseudorange that are on the order of the gravitational radius of the Earth.
\end{abstract}


\maketitle

\section{Introduction}

Clock synchronization algorithms play a key role in applications such as
communication, message encryption, and navigation. Recently, there have been
a number of proposed clock synchronization algorithms based on a quantum
information approach \cite%
{Chuang2000,Jozsa2000,Preskill2000,Giovannetti2001,ShihPRL2003,Burt2001,Jozsa2000a,Yurtsever2000}%
. \ These discussions have been based mostly on non-relativistic quantum
mechanics. In many applications, however, clock synchronization must be
performed between two nodes that are in relative motion, such as a jet
aircraft and the ground, or between two nodes that are at different
gravitational potentials, such as a satellite and the ground, or even two
satellites at different altitudes. In such cases, a correct treatment of
quantum clock synchronization must include relativistic effects from the
start. The well-known tension between quantum mechanics and relativity
theory makes this a difficult task. It is clear that the concepts of
measurement, and transformation of measurable quantities under Lorentz
transformations, play a key role in the problem of clock synchronization for
both quantum and classical schemes. However, the transformation of
measurable quantities has not been discussed even for the case of classical
clock synchronization. \ In this article, we analyze in some detail the
relativity of clock synchronization in the Global Positioning System (GPS),
which is based on a classical synchronization scheme. In particular, we
describe the transformation properties of the measured quantity called
pseudorange in the GPS \cite{ParkinsonGPSReview,Kaplan96,Hofmann-Wellenhof93}%
. We hope that clarifying the transformation properties of measurable
quantities in classical clock synchronization will provide some useful
insight into the problem of quantum clock synchronization.

The Global Positioning System (GPS) is a U.S. military constellation of
satellites used for time keeping, and for navigation of land, air and sea~%
\cite{ParkinsonGPSReview,Kaplan96,Hofmann-Wellenhof93,GLONASS}. Recently,
two papers have analyzed the system of space-time coordinates that is used
in the GPS~\cite{Rovelli2002,Blagojevic2002}. In this paper, we address a
different but related aspect: the transformation of GPS pseudorange
measurements. \ \ A remarkable aspect of the GPS is that a receiver need not
be stationary with respect to the Earth's surface to obtain accurate time
and position. A ship, a jet aircraft, or a low-Earth-orbit satellite can
each compute accurate time and position, even though they have different
velocities. This feature of the GPS is a consequence of two aspects: the
signal structure of the satellite broadcasts and the special type of
measurement that a GPS receiver makes. The GPS satellite signals set up an
invariant grid of 3-dimensional space-time hypersurfaces (light-cones). Each
hypersurface is uniquely marked by the satellite that generated the
hypersurface and by the space-time coordinates of \ the event of generation
of the light-cone \cite{Rovelli2002,Blagojevic2002}. By measuring the
pseudorange to four satellites, the GPS receiver essentially determines its
position by identifying the four hypersurfaces that it intersects. The
pseudorange measurement is independent of receiver motion, up to an additive
constant. For this reason, the measured pseudorange may be called a Lorentz
pseudo-invariant. Below, we will see that the pseudorange is actually a
two-point scalar under generalized coordinate transformations. The
transformation properties of the pseudorange is a key element of the GPS,
yet this subject has only briefly been mentioned in the literature~\cite%
{KrallBahder2001}.

In order to present a coherent description of the transformation properties
of the pseudorange, we must deal with the nature of the broadcast GPS
signals, the relativistic effects that impact these signals, as well as the
measurement process itself. Therefore, the outline of this article is as
follows. \ In section II, we present a two-receiver thought experiment to
clarify the concept of invariance of pseudorange measurements. \ Section III
contains a description of the space-time metric\ in the vicinity of the
Earth. Section IV describes the relativistic effects on the GPS\ satellite
clocks and on the observed signals, using the metric in section III. Section
V discusses the nature of the GPS broadcast signals, which are used to set
up the space-time grid. Section VI discusses the pseudorange measurement
process, using a mechanical analogue for a GPS\ receiver. \ The
transformation properties of the pseudorange are obtained in section VI. \
Section VII contains a brief discussion of navigation and time transfer
using GPS signals. \ A correction is derived to the conventional flat
space-time model of pseudorange. \ Section IX\ contains a summary and
comments.

\section{Two Receiver Experiment}

\begin{figure}
\includegraphics{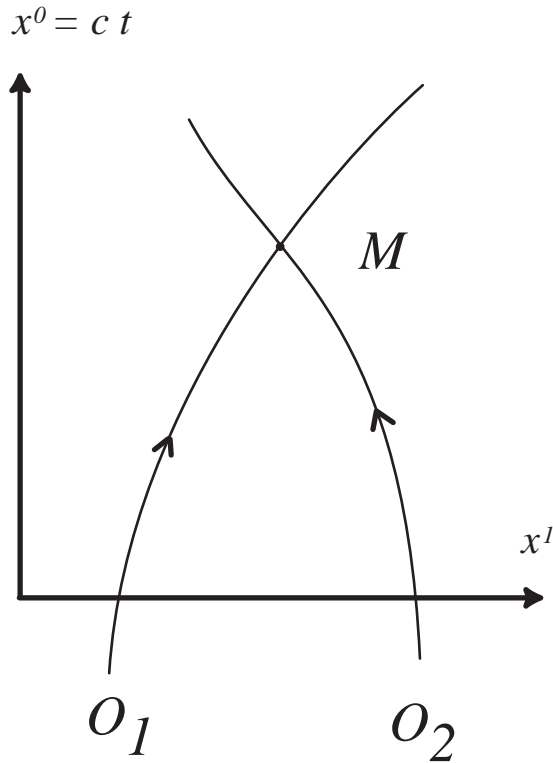}
\caption{\label{TwoReceiverExperiment} The world lines of two
receivers, $O_{1}$ and $O_{2}$, are shown. The receivers are
tracking the same four GPS satellites. At event M the receivers
coincide in space-time but have different velocities.}
\end{figure}

In order to make clear the concept of transformation of pseudorange
measurements we offer a simple thought experiment. Consider two identical
GPS receivers that are in relative motion. For example, one receiver is
stationary with respect to the Earth's surface and the other is on a jet
aircraft travelling at 1000 km/hour. Each GPS receiver carries an identical
copy of the software that is used to compute receiver position and time.
Assume that each receiver is tracking the same four GPS satellites. See Fig.~%
\ref{TwoReceiverExperiment}. Assume the two receivers' world lines cross at
an event $M$ in space-time. Does each receiver compute the same spatial
position and time for the coincident event $M$? From a physics stand point,
the GPS satellites orbit the Earth at approximately 8.37 km/s, and there are
large Doppler frequency shifts due to satellite and receiver motion~\cite%
{RelativisticEffects}. In the GPS, the actual computation of
receiver position and time depends on the space-time coordinates
of the signal emission event (at the satellite), $x_{s}^{i}$, and
reception event (at the receiver), $x_{o}^{i}$. In an
Earth-centered inertial (ECI) frame, these coordinates are
$x_{s}^{i}$ and $x_{o}^{i}$, however in the receiver's comoving
frame the emission and reception events have different
coordinates, say $x_{s}^{\prime i}$ and $x_{o}^{\prime i}$. The
two sets of coordinates, $(x_{s}^{i}$, $x_{o}^{i})$ and
$(x_{s}^{\prime i}$, $x_{o}^{\prime i})$, for $ i=0,1,2,3,$ are
related by a Lorentz transformation, which depends on receiver
velocity with respect to the ECI frame. The actual electromagnetic
field is different in each receiver's comoving frame, so there is
different input information into each identical measuring device
(receiver computer program). How can the two receivers compute the
same spatial position and time from different input information?
 The key concepts are the space-time grid that is created by
GPS satellites and the transformation properties of the measurable
quantity in the GPS, which is called the pseudorange. These two
themes are developed in the following sections.

\section{Choice of Metric}

In order to discuss relativistic effects, a reference frame or system of
coordinates must be chosen. In the weak field limit, the metric of
space-time in the vicinity of the Earth is of the form~\cite%
{LLClassicalFields}

\begin{equation}
-ds^{2}=g_{ij}dx^{i}dx^{j}=-(1+\frac{2}{c^{2}}V)(d{\bar{x}}^{0})^{2}+(1-%
\frac{2}{c^{2}}V)\left[ (dx^{1})^{2}+(dx^{2})^{2}+(dx^{3})^{2}\right]
\label{EarthMetric}
\end{equation}
where $V$ is the Newtonian gravitational potential and $({\bar{x}}%
^{0},x^{1},x^{2},x^{3})$, are the coordinates. The frame of reference in
Eq.~(\ref{EarthMetric}) can be thought of as an Earth-centered inertial
(ECI) frame. We neglect small off-diagonal terms $g_{0\alpha}$, $%
\alpha=1,2,3 $, due to the rotation of the Earth.

In general relativity, the coordinates are mathematical entities that are
never directly observed. However, it is useful to choose the coordinates in
some physically meaningful way. The coordinates ${\bar{x}}%
^{0},x^{1},x^{2},x^{3}$ are geocentric coordinates, where $x^{3}$ coincides
with the Earth's axis of rotation and increasing positive values point to
North. The Earth is modelled as an oblate spheroid with Newtonian potential
given by~\cite{Caputo1967}
\begin{equation}
V(r,\theta)=-\frac{GM}{r}\left[ 1-J_{2}\left( \frac{R}{r}\right)
^{2}P_{2}(\cos(\theta))\right]  \label{EarthPotential}
\end{equation}
where, $r^{2}=(x^{1})^{2}+(x^{2})^{2}+(x^{3})^{2}$ and $\theta$ is the polar
angle measured from the $x^{3}$ axis. In Eq.~(\ref{EarthPotential}), $G$ is
Newton's gravitational constant, $M$ is the mass of the earth, $%
P_{2}(x)=(3x^{2}-1)/2$ is the second Legendre polynomial, $R$ is the Earth's
equatorial radius, and $J_{2}$ is the Earth's quadrupole moment, whose value
is approximately $J_{2}=1.08\times10^{-3}$, see Table I. The metric in Eq.~(%
\ref{EarthMetric}) is the solution of the linearized Einstein field
equations~\cite{LLClassicalFields}. In Eq.~(\ref{EarthMetric}), the
coordinate time ${\bar{x}}^{0}$ has no simple relation to the time kept by
ideal clocks on the surface of the Earth.

The coordinate time can be given a simple interpretation by transforming the
metric in Eq.~(\ref{EarthMetric}) to rotating Earth-centered Earth-fixed
(ECEF) coordinates $y^{i}$, using the transformation
\begin{eqnarray}
{\bar{x}}^{0}  & = &{\bar{y}}^{0}  \nonumber \\
x^{1} & = & \cos(\frac{\omega}{c}{\bar{y}}^{0})\,y^{1}-\sin(\frac{\omega}{c}{%
\bar{y}}^{0})\,y^{2}  \nonumber \\
x^{2} & = & \sin(\frac{\omega}{c}{\bar{y}}^{0})\,y^{1}+\cos(\frac{\omega}{c}{%
\bar{y}}^{0})\,y^{2} \nonumber \\
x^{3}  & = & y^{3}  \label{CoordinateTransformation1}
\end{eqnarray}
Note that the coordinate time in the rotating frame, ${\bar{y}}^{0}$, is
equal to the coordinate time in the ECI frame ${\bar{x}}^{0}$. In these ECEF
rotating coordinates, the metric is given by

\begin{eqnarray}
-ds^{2} & =-\left[ 1+\frac{2V}{c^{2}}-\frac{\Omega^{2}}{c^{2}}\left[
(y^{1})^{2}+(y^{2})^{2}\right] +\frac{2V}{c^{2}}\frac{\Omega^{2}}{c^{2}}%
\left[ (y^{1})^{2}+(y^{2})^{2}\right] \right] (d{\bar{y}}^{0})^{2}  \nonumber \\
& +(1-\frac{2V}{c^{2}})\left[ 2\frac{\Omega}{c}(y^{1}dy^{2}-y^{2}dy^{1})d{%
\bar{y}}^{0}+(dy^{1})^{2}+(dy^{2})^{2}+(dy^{3})^{2}\right]
\label{RotatingCoordMetric}
\end{eqnarray}
From Eq.~(\ref{RotatingCoordMetric}), we see that there exist geopotential
surfaces, $\phi(y^{1},y^{2},y^{3})=c$, where\ $c$ is a constant,
\begin{equation}
\phi(y^{1},y^{2},y^{3})=V-\frac{1}{2}\Omega^{2}\left(
(y^{1})^{2}+(y^{2})^{2}\right)  \label{geoPotDef}
\end{equation}
where stationary clocks in the ECEF\ frame (that satisfy $dy^{\alpha}=0$)
have the same rate of proper time $d\tau=ds/c$\ with respect to coordinate
time ${\bar{y}}^{0}$~\cite{AshbyInParkinsonGPSReview}. In other words, ideal
clocks located at the same value of geopotential $\phi$, have the same rate
with respect to coordinate time $y^{0}$. We have neglected the small
cross-term $2V\Omega^{2}R^{2}/c^{4}\sim10^{-21}$.

Using the observation that clocks at a constant value of geopotential run at
the same rate, it is advantageous to define~\cite{AshbyInParkinsonGPSReview}
the new coordinate time $t$
\begin{equation}
ct=x^{0}=y^{0}=\left( 1+\frac{\phi_{o}}{c^{2}}\right) {\bar{y}}^{0}=\left( 1+%
\frac{\phi_{o}}{c^{2}}\right) {\bar{x}}^{0}  \label{NewCoordTime}
\end{equation}
where $\phi_{o}$ is the value of the geopotential $\phi$ on the Earth's
equator, at $\theta=\pi/2$\ and $r=R$:
\begin{equation}
\phi_{o}=-\frac{GM}{R}(1+\frac{1}{2}J_{2})-\frac{1}{2}\Omega^{2}R^{2}
\label{geopotential}
\end{equation}
For the values of the constants in Table I, the dimensionless magnitude of
this term is $\phi_{o}/c^{2}=-6.96928\times10^{-10}$. Using the
transformation in Eq.~(\ref{NewCoordTime}), the metric in Eq.~(\ref%
{RotatingCoordMetric}) becomes

\begin{eqnarray}
-ds^{2} & =-\left[ 1+\frac{2}{c^{2}}\left( \phi-\phi_{o}\right) \right]
(dy^{0})^{2}+\left( 1-\frac{2V+\phi_{o}}{c^{2}}\right) 2\frac{\Omega}{c}%
(y^{1}dy^{2}-y^{2}dy^{1})dy^{0} \nonumber \\
& +(1-\frac{2V}{c^{2}})\left[
(dy^{1})^{2}+(dy^{2})^{2}+(dy^{3})^{2}\right]
\label{RotatingCoordMetric2}
\end{eqnarray}
Eq.~(\ref{RotatingCoordMetric2}) gives the space-time metric in
ECEF rotating coordinates $y^{i}$. Note that an ideal clock that
is stationary in ECEF coordinates (with $dy^{\alpha}=0$), has
proper time
\begin{equation}
d\tau=ds/c=\frac{1}{c}\left[ 1+\frac{2}{c^{2}}\left( \phi-\phi_{o}\right) %
\right] ^{1/2}\,dy^{0}
\end{equation}
When this clock is located on the geoid, then $\phi-\phi_{o}=0$, and $%
d\tau=dy^{0}/c$, so this ideal clock keeps coordinate time, $x^{0}=y^{0}$.
Hence a good hardware clock that is on the geoid, and stationary with
respect to the rotating Earth, can be used as a reference clock to keep
coordinate time. Note that by Eq. (\ref{NewCoordTime}) the coordinate time
in rotating ECEF coordinates is the same as coordinate time in ECI
coordinates, therefore, the same clock keeps coordinate time in the ECI
frame, $x^{0}$, and coordinate time in the ECEF frame, $y^{0}$.

Using the coordinate time transformation in Eq.~(\ref{NewCoordTime}), the
metric in Eq.~(\ref{EarthMetric}) becomes

\begin{equation}
-ds^{2}=g_{ij}dx^{i}dx^{j}=-\left[ 1+\frac{2}{c^{2}}\left( V-\phi
_{o}\right) \right] (dx^{0})^{2}+\left( 1-\frac{2}{c^{2}}V\right) \left[
(dx^{1})^{2}+(dx^{2})^{2}+(dx^{3})^{2}\right]  \label{EarthMetricECI}
\end{equation}
Equation~(\ref{EarthMetricECI}) gives the metric in ECI coordinates. The
coordinate time that enters into the metric, $x^{0}$, is the time kept by
ideal clocks on the geoid. This result was the goal of the time
transformation given in Eq.~(\ref{NewCoordTime}). Note however, that in the
ECI frame metric in Eq.~(\ref{EarthMetricECI}), the proper time interval $ds$
on a stationary clock in ECI coordinates (with $dx^{\alpha}=0$), is not
equal to coordinate time interval $dx^{0}$ because in general $V\neq\phi_{o}$%
. \ The ECI coordinate metric, given in Eq.~(\ref{EarthMetricECI}), is
useful for computing the proper time $d\tau=ds/c$ elapsed on-board a
satellite clock, in terms of elapsed coordinate time.

\section{Relativistic Effects in GPS}

\subsection{Satellite Oscillator Frequency Offset}

The clocks in GPS satellites are at a higher gravitational potential than
the clocks on Earth. As observed on the Earth, this difference in the
gravitational potential causes the oscillators of GPS atomic clocks to
appear to run fast, by fractional frequency~\cite{LLClassicalFields}
\begin{equation}
\frac{\Delta\omega}{\omega}=\frac{\phi_{1}-\phi_{2}}{c^{2}}  \label{doppler}
\end{equation}
where the approximate gravitational potential $\phi=-GM/r$, and $r$ is the
distance from the center of the Earth. Here, $\phi_{1}$ and $\phi_{2}$ are
the potentials at the satellite and on the earth surface, respectively. For
GPS, $\Delta\omega/\omega\approx5.28\times10^{-10}$. This effect causes the
satellites clocks to run fast by 45 $\mu$s per day. This is often called a
gravitational red shift, but actually, it is a blue shift (toward higher
frequencies).

In addition to the gravitational frequency shift, the GPS satellites are
moving. Consequently, as observed in the ECI frame, satellite oscillators
exhibit time dilation due to their velocity $v/c\approx8.37$ km/s. The time
dilation effect in special relativity is given by
\begin{equation}
\Delta t=\gamma\Delta t^{\prime}  \label{timeDilation}
\end{equation}
where $\Delta t$ is the time interval in the ECI frame and $\Delta
t^{\prime} $ is the proper time of a clock moving at speed $v$ in the ECI
frame. For GPS satellites, $\gamma\approx1-v^{2}/(2c^{2})\approx8.33%
\times10^{-11}$. With respect to coordinate time in the ECI frame, the time
dilation effect makes the satellite clocks appear to run slow by
approximately 37 $\mu$s per day.

The typical GPS atomic clock stability is 1 part in $10^{-13}$, so
the effect of time dilation and gravitational red shift are each
about $10^{3}$ times larger, and therefore, both effects must must
be taken into account. The net effect of time dilation and
gravitational red shift is that the atomic clocks would run fast
by 38 $\mu$s per day ( = 45 $\mu$s - 7 $\mu$s). This is a huge
effect, which can be measured by the fact that 38 $\mu$s
corresponds to a range error of 38,000 feet per day!

The actual value of the combined effect of the gravitational potential and
time dilation on the frequency of the satellite oscillator is computed using
the metric in Eq.~(\ref{EarthMetricECI}). During a coordinate time $dx^{0}$,
the satellite moves a spatial distance $dx^{\alpha}$, $\alpha=1,2,3$. The
proper time elapsed on the satellite clock, $d\tau_{s}=ds/c$, is related to
elapsed coordinate time, $dx^{0}$, by
\begin{equation}
ds=c\,d\tau_{s}=\left[ 1+\frac{2}{c^{2}}(V_{s}-\phi_{o})-(1-\frac{2V_{s}}{%
c^{2}})v_{s}^{2}\right] ^{1/2}\,\,dx^{0}  \label{OscillatorFrequencyOffset}
\end{equation}
where $V_{s}$ is the Earth's gravitational potential, given in Eq.~(\ref%
{EarthPotential}), evaluated at the position of the satellite and $%
v_{s}^{2}=\delta_{\alpha\beta}\,\frac{dx^{\alpha}}{dx^{0}}\frac{dx^{\beta}}{%
dx^{0}}$ is the square of the satellite velocity divided by $c^{2}$. In Eq.~(%
\ref{OscillatorFrequencyOffset}), we take the Earth's quadrupole potential
to be zero, $J_{2}=0$, which allows a circular orbit for the satellite and
makes $d\tau_{s}/dx^{0}$ independent of the polar angle $\theta$ of the
satellite. Similarly, we approximate the satellite velocity $v_{s}$ by
taking $J_{2}=0$ and assume a circular (zero eccentricity) orbit so that $%
v_{s}^{2}=GM/ac^{2}$. Expanding Eq.~(\ref{OscillatorFrequencyOffset}) to
first order in small quantities, $V_{s}/c^{2}$ and $\phi_{o}/c^{2}$, we have
\begin{equation}
c\frac{d\tau_{s}}{dx^{0}}=1-\frac{3GM}{2c^{2}a}-\frac{\phi_{oo}}{c^{2}}%
=1+\delta  \label{OscillatorFrequencyOffset2}
\end{equation}
where
\begin{equation}
\phi_{oo}=-\frac{GM}{R}-\frac{1}{2}\Omega^{2}R^{2}  \label{phiOO}
\end{equation}
is the value of $\phi_{o}$ when $J_{2}=0$. Equation (\ref%
{OscillatorFrequencyOffset2}) gives an approximate expression for the rate
of proper time on a GPS satellite in circular orbit, with respect to
coordinate time $x^{0}$ in the metric in Eq.~(\ref{EarthMetricECI}). We have
dropped terms $V_{s}v_{s}^{2}/c^{2}\approx10^{-20}$ in Eq.~(\ref%
{OscillatorFrequencyOffset2}), since $v_{s}^{2}\approx10^{-10}$ and $%
V_{s}/c^{2}\approx10^{-10}$.

The ratio of the frequencies of oscillators is inversely proportional to the
clock rates, $\omega_{coord}/\omega_{s}=d\tau_{s}/dx^{0}$, where $%
\omega_{coord}$ is the number of cycles elapsed as counted per unit of
coordinate time $dx^{0}$, and $\omega_{s}$ is the number of cycles elapsed
as counted per unit of proper time $d\tau_{s}$ on-board the satellite.
Therefore, the oscillators of clocks on-board the GPS satellites in circular
orbit (taking $J_{2}=0$) have a frequency shift~\cite%
{AshbyInParkinsonGPSReview}
\begin{equation}
\frac{\Delta\omega}{\omega_{s}}=\frac{\omega_{coord}-\omega_{s}}{\omega_{s}}%
=-\frac{3GM}{2c^{2}a}-\frac{\phi_{oo}}{c^{2}}\equiv\delta_{0}
\label{OscillatorFrequencyOffset3}
\end{equation}
Using the value of JGM-2 constants~\cite{JGM-2constants}
\begin{equation}
\delta_{0}=+4.460963\times10^{-10}  \label{freqOffset}
\end{equation}
If the clocks on-board GPS satellites were allowed to run freely, then Eq.~(%
\ref{OscillatorFrequencyOffset2}) shows that during one day of
elapsed coordinate time $\Delta x^{0}$, a satellite clock would
gain $c(d\tau _{s}/dx^{0}-1)\times 24$ Hours. \ Nominally, the GPS
system is designed to transmit the digital P-code at 10.23 MHz.
However, Eq.~(\ref{OscillatorFrequencyOffset3}) shows that if the
oscillator in the satellite were set to $\omega_{s}/2\pi=10.23$
MHz,
then this code would appear to have a higher frequency, $\omega_{coord}=(1+%
\delta )\omega_{s}$, as measured with respect to a clock that keeps
coordinate time in the ECI frame. If a clock is stationary in the ECEF
frame, and is located on the Earth's geoid, it keeps coordinate time $x^{0}$%
, and therefore the satellite clocks would appear to run fast to a GPS user
on the geoid. Consequently, in the GPS, the technical specifications for the
satellite clocks include a frequency \textquotedblleft factory offset" that
is applied prior to launch of the satellites. The actual (angular) frequency
of P-code that is broadcast by the satellite clock is~\cite%
{SpilkerInParkinsonGPSReview}
\begin{equation}
\omega_{s}=2\pi(1-\delta_{0})f_{0}  \label{FactoryOffset}
\end{equation}
where $\delta_{0}=+4.46\times10^{-10}$ and $f_{0}=10.23$ MHz are fixed GPS
constants. With this frequency correction applied, the satellite clocks
approximately keep coordinate time $x^{0}$ in the ECI frame.

However, satellite orbits cannot be made perfectly circular, so
the GPS clocks are in a slightly eccentric orbit. This
eccentricity of the orbit contributes an additional well-known
effect on the time of the satellite clocks: the satellite clocks
periodically speed up and slow down with respect to coordinate
time $x^{0}$. This effect is often called the \textquotedblleft e
Sin E effect" and depends on the position of the satellite in its
orbit. This clock correction is made in the GPS receiver in
software, and is given by~\cite{ICD-GPS-200}
\begin{equation}
\Delta t_{r}=\frac{2}{c^{2}}\sqrt{GMa}\,e\,\sin E  \label{eSinE}
\end{equation}
where $a$ is the semimajor axis of the satellite, $e$ is the orbital
eccentricity, and $E$ is the eccentric anomaly along the orbit. For a
typical upper limit of $e=0.01$, $\Delta t_{r}\approx23ns$.

\subsection{Observed Frequency Shift}

Equation~(\ref{FactoryOffset}) gives the frequency offset that is
applied to a GPS satellite clock/oscillator so that it
approximately keeps coordinate time $x^{0}$, with respect to a
reference oscillator that is stationary in the ECEF frame and
located on the geoid. Most users of GPS do not satisfy these two
criteria. For example, a user that is stationary in the ECEF frame
on the geoid is moving in the ECI frame, and a user in an aircraft
is above the geoid. Therefore, at any given time, a user typically
sees a frequency shift that is different from the
\textquotedblleft factory offset", given in
Eq.~(\ref{FactoryOffset}). The actual measured frequency shift of
the satellite signal depends on user and satellite  positions in
the gravitational field of the Earth, and also on user and
satellite velocities (not just user-satellite velocity
differences). The reason that the frequency shift depends
separately on user and satellite velocities is, of course, that
space-time is not homogeneous (space-time is not flat) because of
the Earth's gravitational field.
\begin{table}[ptb]
\caption{The numerical constants and values used in the calculations. Values
taken from Ref. \protect\cite{ParkinsonGPSReview} and \protect\cite%
{JGM-2constants}. }
\label{tab:NumericalConstants}%
\begin{ruledtabular}
\begin{tabular}{clll}
Symbol   & Definition       &  Value     &  Units \\ \\
\hline
$G M$    & Gravitational constant time Earth's Mass      &  3.986004415$\times 10^{14}$   (JGM-2)             &  m$^3$/s$^3$ \\
$c$      &  vacuum speed of light                        &   2.99792458$\times 10^{8}$    (exact definition)             &  m/s  \\
$\Omega$ & angular velocity of Earth rotation            &  7.2921151467$\times 10^{-5}$  (WGS-84)            &   radian/s  \\
$R$      & mean Earth radius at Equator                  &  6.3781363$\times 10^6$        (JGM-2)              &       m \\
$J_2$    & Earth's quadrupole moment                     &  1.0826269$\times$10$^{-3}$    (JGM-2)              &  1   \\
$a$      & GPS orbit semimajor axis                      &  26561.75$\times 10^3$          Ref.~\cite{AshbyInParkinsonGPSReview}                      &  m    \\
$v_s / c= \sqrt{G M/a} \, / \, c$   & GSP satellite velocity / c  &  1.29217$\times 10^{-5}$                               &  1  \\
$G M/R c^2$  & gravitational potential scale                          &  6.95348$\times 10^{-10}$                             &  1 \\
\end{tabular}
\end{ruledtabular}
\end{table}

Consider a satellite moving at velocity $\mathbf{v}_{s}$ at gravitational
potential $V_{s}$. At event $S$, the satellite transmits an electromagnetic
signal of proper frequency $\omega_{s}$, as measured with respect to a
calibrated oscillator on-board the satellite. An observer at event $O$ is at
gravitational potential $V_{o}$ and has a velocity $\mathbf{v}_{o}$. The
observer measures a signal having frequency $\omega_{o}$, which is different
from $\omega_{s}$ because he is in motion and at a different gravitational
potential than the satellite. The emission event $S$ and observer reception
event $O$ are connected by a null geodesic. The quantity $%
\omega_{o}/\omega_{s}-1$ is a 2-point scalar because it depends on two
space-time points, $S$ and $O$. A 2-point scalar transforms as a true scalar
under separate transformations of coordinates at point $S$ and at point $O$~%
\cite{Synge1960}. Using the metric in Eq.~(\ref{EarthMetricECI}), a detailed
calculation gives~(see Appendix B)

\begin{eqnarray}
\frac{\omega_{o}}{\omega_{s}}-1 & =(\mathbf{v}_{s}-\mathbf{v}_{o})\cdot%
\mathbf{n}\,\,\left[ 1+\frac{V_{s}-V_{o}}{c^{2}}+\frac{1}{2}%
(v_{o}^{2}-v_{s}^{2})+(\mathbf{v}_{s}\cdot\mathbf{\hat{n}})+(\mathbf{v}%
_{s}\cdot\mathbf{\hat{n}})^{2}-\alpha_{1}-\beta_{1}\right]
\nonumber  \\
& +\frac{V_{s}-V_{o}}{c^{2}}+\frac{1}{2}(v_{o}^{2}-v_{s}^{2})+\frac {2GM}{%
c^{2}r}\left[ h(\mathbf{r}_{o},\mathbf{r}_{s},\mathbf{v}_{o},\mathbf{v}%
_{s})+h(\mathbf{r}_{s},\mathbf{r}_{o},\mathbf{v}_{s},\mathbf{v}_{o})\right]
\label{observedDoppler}
\end{eqnarray}
where $\mathbf{r}_{o}$ and $\mathbf{r}_{s}$ are positions, and $\mathbf{v}%
_{o}$ and $\mathbf{v}_{s}$ are velocities (divided by $c$), of observer and
satellite (source), respectively. In Eq.~(\ref{observedDoppler}), $r=|%
\mathbf{r}_{o}-\mathbf{r}_{s}|$, $v_{o}^{2}=\mathbf{v}_{o}\cdot \mathbf{v}%
_{o}$, $v_{s}^{2}=\mathbf{v}_{s}\cdot\mathbf{v}_{s}$, and $V_{s}$ and $V_{o}$
are the Earth's gravitational potential given in Eq.~(\ref{EarthPotential}),
evaluated at emission event $S$ and reception event $O$, respectively. The
unit vector $\mathbf{\hat{n}}$ connects the emission and reception events, $%
S $\ and\ $O$:
\begin{equation}
\mathbf{\hat{n}}=\frac{\mathbf{r}_{o}-\mathbf{r}_{s}}{|\mathbf{r}_{o}-%
\mathbf{r}_{s}|}  \label{unitVectorN}
\end{equation}
The dimensionless quantities $\alpha_{1}=GM/c^{2}R$ and $\beta_{1}=\Omega
^{2}R^{2}/2c^{2}$. The last term in the right hand side of Eq.~(\ref%
{observedDoppler}) is given by the sum of two terms, which are related by
interchanges of subscripts \textquotedblleft s" and \textquotedblleft o".
The scalar function $h(\mathbf{r}_{o},\mathbf{r}_{s},\mathbf{v}_{o},\mathbf{v%
}_{s})$ is given by

\begin{equation}
h(\mathbf{r}_{o},\mathbf{r}_{s},\mathbf{v}_{o},\mathbf{v}_{s})=\frac {1}{1-(%
\mathbf{\hat{n}}\cdot\mathbf{\hat{r}}_{o})^{2}}\left\{ \left[ \frac{r_{o}}{r}%
-\frac{\mathbf{r}_{o}\cdot\mathbf{r}_{s}}{r\,r_{o}}\right] \mathbf{\hat{n}}%
\cdot(\mathbf{v}_{o}-\mathbf{v}_{s})-\frac{\mathbf{r}_{o}}{r_{o}}\cdot(%
\mathbf{v}_{o}-\mathbf{v}_{s})+\frac{\mathbf{r}_{s}\cdot\mathbf{v}_{o}}{r_{o}%
}-\frac{(\mathbf{r}_{o}\cdot\mathbf{r}_{s})(\mathbf{r}_{o}\cdot\mathbf{v}%
_{o})}{r_{o}^{3}}\right\}  \label{hFunction}
\end{equation}
For a GPS user on the surface of the Earth, $(V_{s}-V_{o})/c^{2}\approx5.3%
\times10^{-10}$, and the GPS satellite velocity (divided by $c$) is $%
v_{s}\approx1.3\times10^{-5}$. We also have that $\alpha_{1}\approx
6.9\times10^{-10}$ and $\beta_{1}\approx1.2\times10^{-12}$. Therefore, in
the derivation of Eq.~(\ref{observedDoppler}), we have taken $v_{s}=O(1)$, $%
V_{s}/c^{2}=O(2)$, $V_{o}/c^{2}=O(2)$, $\alpha_{1}=O(2)$, and
$\beta _{1}=O(2)$, where $O(1)\sim10^{-5}$, and I have dropped
terms $O(4)\sim 10^{-20}$.

The measured frequency is due to three types of terms. First, there is a
special relativistic Doppler effect that depends on the relative velocity of
satellite and observer, contained in the terms proportional to $(\mathbf{v}%
_{s}-\mathbf{v}_{o})\cdot\mathbf{n}$, and the term $\frac{1}{2}%
(v_{o}^{2}-v_{s}^{2})$, which comes from expanding the special relativistic $%
\gamma $-factors for satellite and observer. Next, there is a frequency
shift due to the difference in gravitational potential of the observer and
satellite, which is given by the stand alone term ($V_{s}-V_{o})/c^{2}$.
Finally, there are cross terms that depend on products of satellite and
observer velocities and the Earth's mass $M$.

As described in Eq.~(\ref{FactoryOffset}), the GPS satellites have a
built-in \textquotedblleft factory offset\textquotedblright, $%
\delta_{0}=4.46\times 10^{-10}$. From Eq.~(\ref{observedDoppler}),
we see that the observed frequency shift of the satellite signal
due to gravitational potential differences, $(V_{s}-V_{o})/c^{2}$
is on the order of 10$^{-10}$ and can vary depending on altitude
of the observer. In addition to this frequency shift, there is a
Doppler frequency shift that is much larger. For example, for a
jet aircraft travelling along the equator at approximately
1000 km/hour with respect to the Earth's surface, the velocity (fraction of $%
c$) with respect to the ECI coordinates can be on the order $%
v_{o}\sim2\times10^{-6}$, which is a factor of $5\times10^{3}$ larger than
the built-in \textquotedblleft factory offset\textquotedblright. Similarly,
for an observer on-board a low-Earth orbit satellite at altitude 1000 km,
whose orbit is in the plane of the equator, we have $|\mathbf{v}_{s}-\mathbf{%
v}_{o}|\sim10^{-5}$. This frequency shift is on the order of
10$^{4}$ times larger than the built-in \textquotedblleft factory
offset\textquotedblright. Of course the factory offset is applied
to cancel out a secular effect--an effect that leads to a
constantly increasing discrepancy in time between satellite and
coordinate time clocks, while the Doppler (motional) effect has a
more complicated time dependence. The point here is that the
frequency of GPS signals, as seen by an observer, has huge
frequency shifts due to observer motion. Below, we describe how
these frequency shifts are essentially removed--so that the actual
measurements made by a GPS receiver are independent of the
velocity of the observer (GPS receiver).

\section{GPS Signals: The Space-Time Grid}

The GPS satellites broadcast electromagnetic signals that set up a geometric
space-time grid \cite{Kheyfets1991}. Users of the GPS that receive four
satellite signals can uniquely identify their position in the space-time
grid. This grid is created by discontinuities in the amplitude of the
broadcast electromagnetic field.

\begin{figure}
\includegraphics{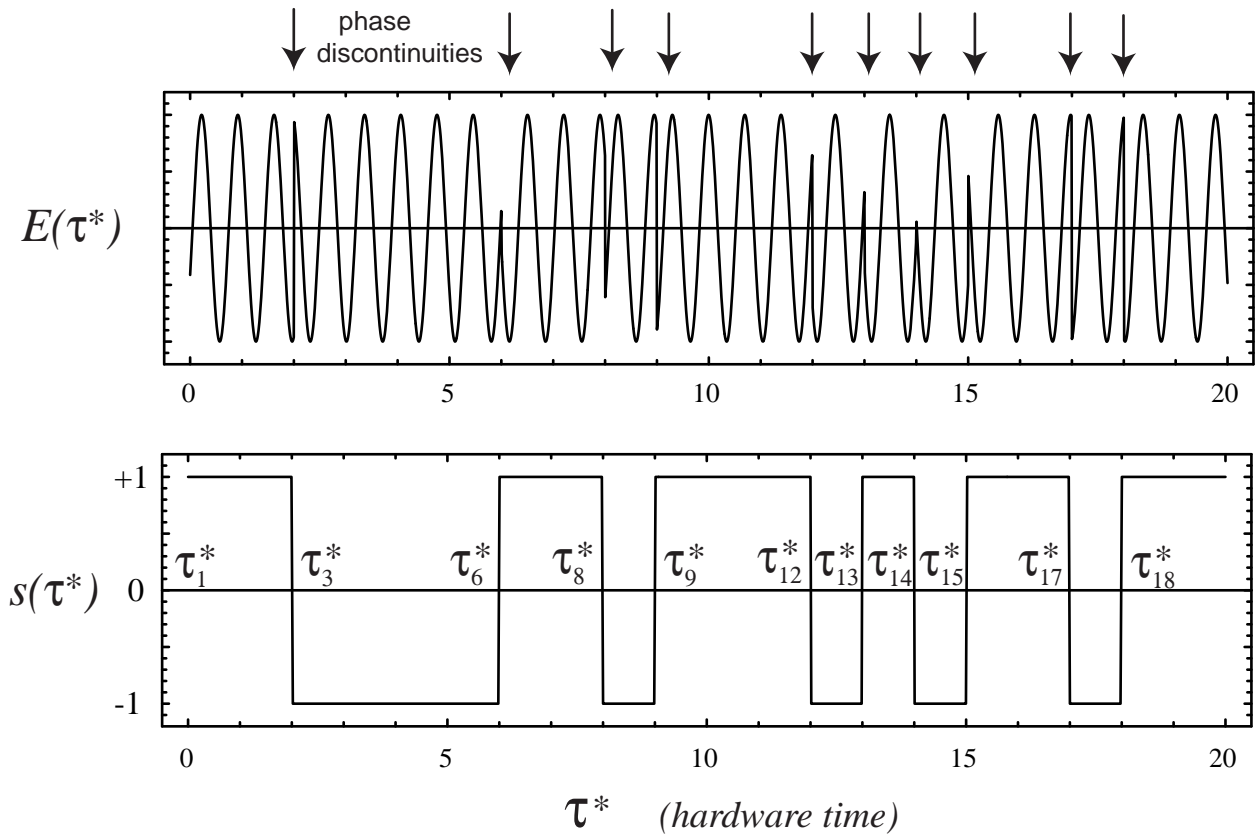}
\caption{\label{SatelliteSignal} Each component of the electric field,
$E(\protect\tau ^{\ast })$, broadcast
by a GPS satellite has possible discontinuities in its amplitude at the
times, $\protect\tau _{1\text{,}}^{\ast }\protect\tau _{2\text{,}}^{\ast }...%
\protect\tau _{N\text{,}}^{\ast }$ where the pseudorandom P-code function $s(%
\protect\tau ^{\ast })$ changes sign. These discontinuities propagate on 3-d
hypersurfaces in space-time and define the space-time grid.}
\end{figure}

All the GPS satellites broadcast on two carrier frequencies in the L-band
centered about: $L_{1}\approx1575.42$ MHz and $L_{2}\approx1227.6$ MHz. The
GPS satellites have helical antennas that are constantly pointed toward the
Earth center. Near the antenna axis (bore-site) the broadcast
electromagnetic radiation field is right circularly polarized \cite%
{KrallBahder2001}. Superimposed on each satellite carrier signal is a unique
code, or sequence of digital bits that identifies the satellite that is
broadcasting. At a given time and spatial position in the comoving frame of
the satellite, one vector component of the electric field can be written
approximately as
\begin{equation}
E=D(\tau^{\ast})s(\tau^{\ast})\cos(q\,\omega_{s}\tau^{\ast}+\phi )
\label{ElectricField}
\end{equation}
where $\omega_{s}$ is the P-code transmission frequency given in Eq.~(\ref%
{FactoryOffset}), $\tau^{\ast}$ is the hardware time kept on-board the
satellite by its local clock, and $\phi$ is a phase associated with phase
noise in the signal. The integer $q$ takes values $q=154$ or $q=120$, to
produce the broadcast signal that is transmitted on the two carrier
frequencies, $L_{1}=154\times\omega_{s}/2\pi\approx1575.42$ MHz or $%
L_{2}=120\times\omega_{s}/2\pi\approx1227.6$ MHz. See Fig.~\ref%
{SatelliteSignal}. The $L_{1}$ and $L_{2}$ carrier frequencies are integer
multiples of the code bit transmission rate, $\omega_{s}/2\pi$. We
distinguish between proper time $\tau$ kept by an ideal clock, and hardware
time, $\tau^{\ast}$, which is kept by a clock that is a real physical
device. The function $s(t)$ takes the discrete values $-1$ and $+1$, and
represents the digital P-code values $0$ and $1$ in the pseudorandom code,
which is unique to each satellite \cite{SpilkerInParkinsonGPSReview}. There
are exactly $N(\approx10^{12})$ values in the digital P-code sequence. The
code sequence starts at midnight on Sunday and has a period of exactly one
week: $s(t+T)=s(t)$, where $T$ is one week. Therefore, we can define a
discrete phase $\Phi_{n}$ for the periodic function $s(t)$ by
\begin{equation}
\Phi^{(n)}=2\pi\frac{(n-1)}{N}  \label{CodePhase1}
\end{equation}
where $n=1,2,3,...,N$, is an integer that sequentially labels the bits in
the code $s(t)$.

In Eq. (\ref{ElectricField}), the function $D(\tau ^{\ast })$ is a digital
navigation message that is broadcast at approximately 50 Hz. The message
provides the satellite ephemeris and satellite clock corrections in the form
of two coefficients, $A$ and $B$. The clock corrections essentially provide
the conversion from hardware time $\tau ^{\ast }$ to coordinate time in the
ECI frame, $x^{0}$, in the form $x^{0}=A+B\tau ^{\ast }$. The functions $%
s(\tau ^{\ast })$ and $D(\tau ^{\ast })$ are timed so that changes between 0
and 1 in $s(\tau ^{\ast })$ occur at the same point in time as in $D(\tau
^{\ast })$, i.e., the bit transitions (edges) in $D(\tau ^{\ast })$ align
with those in $s(\tau ^{\ast })$. The function $D(\tau ^{\ast })$ provides
the broadcast ephemeris of each satellite.

\begin{figure}
\includegraphics{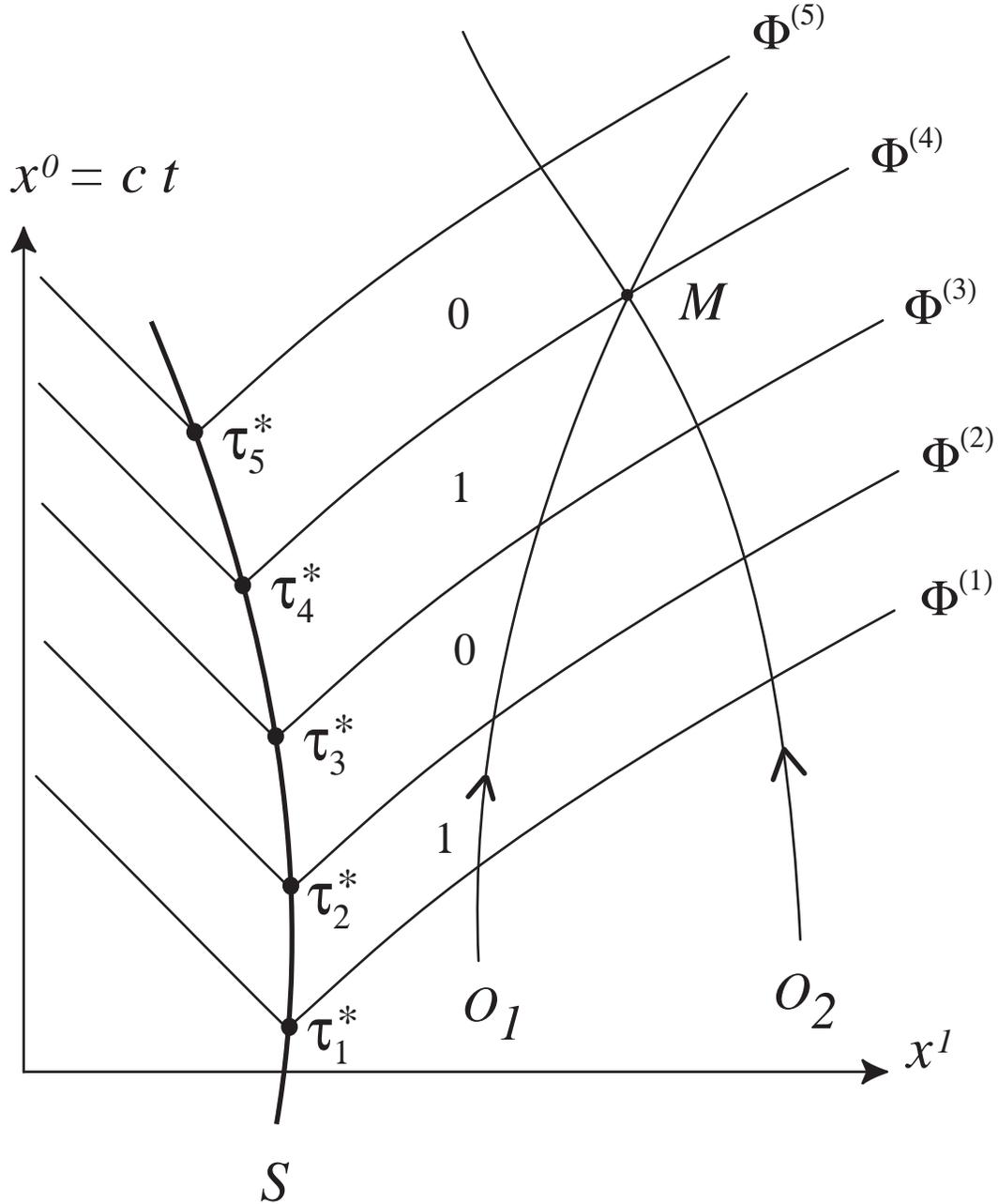}
\caption{\label{SpaceTimeGrid} Discontinuities in the amplitude of
the electromagnetic field  at times, $\tau_{1}^{\ast }\text{,}
\tau_{2}^{\ast }\text{,} \cdots \tau_{N}^{\ast }\text{,}$
propagate on 3-d hypersurfaces in space-time that define the
space-time grid.}
\end{figure}

In the satellite comoving frame, the pseudorandom code, $s(\tau^{\ast})$, is
broadcast at the factory adjusted angular frequency $\omega_{s}$ given in
Eq.~(\ref{FactoryOffset}), as timed by the atomic clock on-board each
satellite. The frequency of the oscillator (clock) on-board the satellite
has been lowered from $f_{0}$ MHz to $\omega_{s}/2\pi$ (see Eq. (\ref%
{FactoryOffset})) so that, as seen from the ECI frame of reference, the
discontinuities in the amplitude of the electromagnetic field occur at time
intervals of approximately $1/f_{0}$ \textit{with respect to coordinate time}
in the metric given in Eq.~(\ref{EarthMetricECI})~\cite{eSinE}. Since the
code $s(\tau^{\ast})$ is timed according to the satellite clock, the
discontinuities in the code can occur at hardware times $\tau_{n}^{\ast}$, $%
n=1,2,3,...,N$, where $n$ sequentially labels the (potential) code edge
discontinuities since midnight Sunday, and $N$ is the number of code bits in
the $s(\tau^{\ast})$ code. All satellites have the same sequence of hardware
times $\tau_{n}^{\ast}$ for possible code-edge-emission events, given by
\begin{equation}
\tau_{n}^{\ast}=\frac{\Phi^{(n)}}{\frac{\omega_{s}}{N}}=\frac{2\pi}{\omega
_{s}}(n-1)  \label{CodeTimeMapping1}
\end{equation}
where $\omega_{s}$ is a constant frequency given in Eq.~(\ref{FactoryOffset}%
) and $\;n=1,2,3,\cdots N$ is the bit number in the P-code sequence $%
s(\tau^{\ast})$. In the ECI frame of reference, the coordinates of the world
line of the antenna phase center for satellite $s$, $x_{s}^{i}(\tau^{\ast})$%
, can be parametrized by the satellite clock's hardware time, $\tau^{\ast}$.
The world line of a satellite, is approximately a geodesic, however, forces
on a satellite, such as solar pressure, and antenna phase center position
differing from the satellite center of mass, lead to an effective world line
that differs from a geodesic. The world line of the antenna phase center is
precisely determined by ground tracking stations. Future points on the world
line are computed and expressed in terms of classical satellite ephemeris
parameters and the $A$ and $B$ clock corrections. This information is
uploaded to the satellite's navigation message, which is transmitted to GPS
users through the digital sequence $D(\tau^{\ast})$.

It is well known that discontinuities in the emitted electromagnetic field
tensor define invariant 3-d hypersurfaces called characteristics \cite%
{Synge1960,Fock1964,DeFeliceClarke1990}. In terms of the world function of
the space-time, for each satellite, these hypersurfaces are given by~\cite%
{Bahder2001}
\begin{equation}
\Omega(x_{s}^{i}(\tau_{n}^{\ast}),x^{j})=0\;\;,\;\;\;\;n=1,2,3,\cdots N
\label{spacetimeGrid}
\end{equation}
where $x_{s}^{i}(\tau^{\ast})$ is the world line of the satellite $s$\
parametrized by satellite hardware time $\tau^{\ast}$, and the 3-d
hypersurfaces are defined by their coordinates $x^{j}$, $j=0,1,2,3$, that
satisfy Eq.~(\ref{spacetimeGrid}) and are on the forward light cone, so that
$x^{0}>x_{s}^{0}(\tau_{n}^{\ast})$. Each hypersurface can be uniquely
specified by $\tau_{n}^{\ast}$, the hardware time of satellite
\textquotedblleft s" and the bit number $n$ associated with the
(discontinuity) new bit in the code $s(\tau^{\ast})$. For flat space-time
with a Minkowski metric, Eq.~(\ref{spacetimeGrid}) reduces to
\begin{equation}
\frac{1}{2}\eta_{ij}\,(x^{i}-x_{s}^{i}(\tau_{n}^{\ast}))\,(x^{j}-x_{s}^{j}(%
\tau_{n}^{\ast}))=0  \label{WorldFunctionFlatSpaceTime}
\end{equation}
where $\eta_{ij}$ is the Minkowski metric with nonzero diagonal components $%
(-1,+1,+1,+1)$, and there is no sum on $n$.

A detailed calculation of the world function entering in Eq.~(\ref%
{spacetimeGrid}) for metric in Eq.~(\ref{EarthMetricECI}) gives~

\begin{equation}
\Omega(x_{1}^{i},x_{2}^{j})=-\frac{1}{2}(\Delta x^{0})^{2}\,\,\left[ 1+\frac{%
2GM}{c^{2}R}+\frac{\Omega^{2}R^{2}}{c^{2}}-\frac{2GM}{c^{2}\,\,|\Delta%
\mathbf{r}|}\,\Lambda(\mathbf{r}_{1},\mathbf{r}_{2})\right] +\frac{1}{2}%
\,(\Delta\mathbf{r})^{2}\left[ 1+\frac{2GM}{c^{2}\,|\Delta \mathbf{r}|}%
\Lambda(\mathbf{r}_{1},\mathbf{r}_{2})\right]  \label{WorldFunction}
\end{equation}
where in Eq.~(\ref{WorldFunction}) we use the following definitions, $%
x_{1}^{i}=(x_{1}^{0},\mathbf{r}_{1})$, $x_{2}^{i}=(x_{2}^{0},\mathbf{r}_{2})$%
, $\Delta x^{0}=x_{1}^{0}-x_{2}^{0}$ and $|\Delta\mathbf{r}|=|\mathbf{r}_{1}-%
\mathbf{r}_{2}|$,
\begin{equation}
\Lambda(\mathbf{r}_{1},\mathbf{r}_{2})=\log\left( \frac{\tan(\frac{\theta
_{1}}{2})}{\tan(\frac{\theta_{2}}{2})}\right)  \label{Lambda12 Function}
\end{equation}
and $\theta_{1}$ and $\theta_{2}$ are defined by
\begin{equation}
\cos\theta_{a}=\frac{\mathbf{r}_{a}\cdot(\mathbf{r}_{2}-\mathbf{r}_{1})}{|%
\mathbf{r}_{a}||\mathbf{r}_{2}-\mathbf{r}_{1}|},\;\;\;\;a=1,2
\label{cosineDef2}
\end{equation}
In Eq.~(\ref{WorldFunction}), we have taken the small parameter $J_{2}=0$.

Each GPS satellite broadcasts a set of 3-d hypersurfaces that form part of
the coordinate grid, given in Eq.~(\ref{spacetimeGrid}). There are
approximately 24 satellites in the GPS constellation, and all the
hypersurfaces from these satellites comprise the GPS space-time grid that is
used to label events in this space-time. The hypersurfaces are spaced
approximately $1/f_{0}\approx$ 97.75 ns in coordinate time and 29.31 m in
space, see metric in Eq.(\ref{EarthMetricECI}). Since the coordinate
hypersurfaces are so far apart in space and time, an event in this
space-time can be specified more accurately by interpolating the times $%
\tau_{n}^{\ast}$ at which discontinuities in electromagnetic field are
actually emitted. For each satellite, $s$, an interpolated hardware time, $%
\tau_{\eta}^{\ast}$, and corresponding interpolated phase of pseudorandom
code, $\Phi$, can be associated with a real number $\eta$
\begin{equation}
\tau_{\eta}^{\ast}=\frac{\Phi}{\frac{\omega_{s}}{N}}=\frac{2\pi}{\omega_{s}}%
(\eta-1)  \label{xiTime}
\end{equation}
where $\omega_{s}$ is given by Eq.~(\ref{FactoryOffset}), $%
\tau_{n}^{\ast}\leq\tau_{\eta}^{\ast}<\tau_{n+1}^{\ast}$ and $n\leq\eta<n+1$%
, see Eq.~(\ref{CodeTimeMapping1}). The real number $\eta$ is the
interpolated value of the code bit integer number $n$. Associated with the
interpolated hardware time, $\tau_{\eta}^{\ast}$, there is a continuous
family of coordinate 3-d hypersurfaces analogous to those in Eq.~(\ref%
{spacetimeGrid}):
\begin{equation}
\Omega(x_{s}^{i}(\tau_{\eta}^{\ast}),x^{j})=0
\label{spacetimeGridInterpolated}
\end{equation}
The continuous parameter $\tau_{\eta}^{\ast}$ labels the 3-d hypersurface
that has code phase $\Phi$ and is defined by coordinates $x^{j}$ that
satisfy Eq.~(\ref{spacetimeGridInterpolated}). On each 3-d hypersurface, the
phase $\Phi$ has the \ value
\begin{equation}
\Phi=\frac{\omega_{s}}{N}\tau_{\eta}^{\ast}  \label{Phase on hypersurface}
\end{equation}
where $\omega_{s}$ is given by Eq. (\ref{FactoryOffset}). The
hardware time $\tau_{\eta}^{\ast}$ of an emission event at the
satellite\ is related to coordinate time $t_{s}=x_{s}^{0}/c$ in
the ECI\ frame of reference by the satellite clock correction
$\Delta\tau_{\eta}^{\ast}$
\begin{equation}
t_{s}=\tau_{\eta}^{\ast}+\Delta\tau_{\eta}^{\ast}
\end{equation}
In terms of coordinate time of emission, the phase broadcast by satellite $s$
is then given by
\begin{equation}
\Phi_{s}(t,\mathbf{r})=\frac{\omega_{s}}{cN}\left( x_{s}^{0}(t,\mathbf{r}%
)-c\Delta\tau_{\eta}^{\ast}\right)
\label{Phase on hypersurface with clock offset}
\end{equation}
where $\omega_{s}\Delta\tau_{\eta}^{\ast}/N$represents a phase
offset due to the fact that satellite clocks keep hardware time,
which is an approximation of coordinate time $t_{s}=x_{s}^{0}/c$.
The broadcast phase function  satisfies the eikonal
equation\cite{Fock1964}
\begin{equation}
g^{ij}\frac{\partial\Phi_{s}}{\partial x^{i}}\frac{\partial\Phi_{s}}{%
\partial x^{j}}=0  \label{EikonalEq}
\end{equation}
where $g^{ij}$ are the contravariant components of the metric given in Eq.~(%
\ref{EarthMetricECI}) (no sum on $s$). The phase function $%
\Phi_{s}$ depends on the invariant world line of the satellite, $%
x_{s}^{i}(\tau^{\ast})$. \ From Eq.(\ref{EikonalEq}) it is clear that under
Lorentz transformations, or generalized coordinate transformations, the
phase $\Phi_{s}(t,\mathbf{r)}$ transforms as a scalar. The form of Eq.(\ref%
{EikonalEq}) shows that the wave vector associated with the phase $\Phi_{s}$%
, $\kappa_{i}=\partial\Phi_{s}/\partial x^{i}$, is a null vector. \ The wave
vector $\kappa_{i}$ can be related to the covariant derivative of the world
function of space-time, $\Omega(T,R)$, between emission event $T=(t_{T},%
\mathbf{r}_{T})$\ and reception event $R=(t,\mathbf{r})$. \ The direction of
the wavevector $\kappa_{i}$ is the same as the direction of the covariant
derivative of the world function, $\Omega_{i_{R}}$, where the derivative is
taken with respect to $R$. \ This can be seen from the identity $%
\Omega=g^{ij}(R)\Omega_{i_{R}}\Omega_{j_{R}}=0$, since the geodesic
connecting points $T=(t_{T},\mathbf{r}_{T})$ and $R=(t,\mathbf{r})$ is null.

A world line is an invariant geometric quantity\cite{Synge1960}.
For the limiting case of a satellite in Minkowski space-time, the
time component $t_{s}$ of the satellite world line in
Eq.(\ref{Phase on
hypersurface with clock offset}) defines a scalar field, $t_{s}=t_{s}(t,\mathbf{r)%
}$, that is given by the implicit equation
\begin{equation}
t_{s}(t,\mathbf{r)}=t-\frac{1}{c}|\mathbf{r}-\mathbf{r}_{s}(t_{s}(t,\mathbf{%
r)})|  \label{FlatSpaceLightCone}
\end{equation}
where $\mathbf{r}_{s}(t_{s})$ is the ephemeris of the satellite. \ It is
easy to check that this particular form for $t_{s}(t,\mathbf{r)}$ in the
phase function $\Phi_{s}(t,\mathbf{r})$\ in Eq.(\ref{Phase on hypersurface
with clock offset}) satisfies the eikonal Eq.(\ref{EikonalEq}).

For the actual case where we take into account the Earth's gravitational
field, the light cone equation is given by $\Omega(T,R)=0$, where $\Omega$
is the world function given in Eq.(\ref{WorldFunction}). \ Writing the world
function in Eq.(\ref{WorldFunction}) in the form%
\begin{equation}
\Omega=-\frac{1}{2}(1+\alpha)(x^{0}-x_{s}^{0})^{2}+\frac{1}{2}(1+\beta )(%
\mathbf{r-r}_{s})^{2}  \label{WorldFunction alpha beta form}
\end{equation}
where $\alpha$ and $\beta$ are small quantities, the scalar phase field $%
\Phi_{s}(\mathbf{r,}t\mathbf{)}$ in Eq.(\ref{Phase on hypersurface with
clock offset}) can be written in terms of the function $x_{s}^{0}(x^{0},%
\mathbf{r})$ that is implicitly given by
\begin{equation}
x_{s}^{0}(x^{0},\mathbf{r})=x^{0}-\left( 1+\frac{1}{2}(\beta-\alpha)\right) |%
\mathbf{r-r}_{s}(x_{s}^{0}(x^{0},\mathbf{r}))|
\label{time field with gravity}
\end{equation}
where we have kept only linear terms in $\alpha$ and $\beta$. \ Note that $%
\alpha$ and $\beta$ are two-point functions that depend on space-time points
$T=(x_{s}^{0},\mathbf{r}_{s})$ and $R=(x^{0},\mathbf{r})$.

An event in this space-time can be uniquely labelled by four real hardware
times, $(\tau_{1}^{\ast},\tau_{2}^{\ast},\tau_{3}^{\ast},\tau_{4}^{\ast})$,
or alternatively by dimensionless real numbers, $(\eta_{1},\eta_{2},%
\eta_{3},\eta_{4})$, where each $\eta_{s}$ gives the continuous code
parameter at emission time for satellite $s$. \ This system of coordinates
has been studied by Synge who called them optical coordinates \cite%
{Synge1960}. More recently, these same coordinates have been called GPS
coordinates and their theoretical properties of have been explored in some
detail\cite{Rovelli2002,Blagojevic2002}.

At any time, a GPS user has more than four satellites in view, so in the
real implementation of the GPS system of coordinates, they are multiple
valued. Since the satellite signals are line-of-sight, a GPS user sees
satellites rise and set on the horizon, and a different set of four
satellites defines the 3-d hypersurfaces. There are currently approximately
24 satellites in the GPS constellation, and 37 code sequence possibilities~%
\cite{SpilkerInParkinsonGPSReview}, so at most, the numbers $%
s=1,2,3,\cdots37 $.

In our discussion, we have neglected atmospheric effects. In practice, for
users of GPS in the Earth's atmosphere, there exist significant propagation
delays as well as frequency dispersive effects. The Earth's troposphere
(atmosphere from the ground to approximately 10 km) causes the same time
delay for both $L_{1}$ and $L_{2}$ frequencies. On the other hand, the
Earth's ionosphere (60 km to 700 km altitude) is dispersive at these
frequencies, due to the presence of free electrons, and hence causes two
different delays for the $L_{1}$ and $L_{2}$ frequencies. A first principles
treatment is possible, following Synge~\cite{Synge1960}. However, in
practice, simpler methods are used for correcting for both of these effects
in GPS receivers~\cite{SpilkerInParkinsonGPSReview}. \ Atmospheric effects
will not be considered further here. However, we note that signal
propagation through the (frequency dispersive) atmosphere leads to a removal
of the discontinuity in the electromagnetic tensor $F_{ij}$. Hence, the
dispersive effects of the atmosphere limit the accuracy of GPS for a ground
user because the code edge is not sharp after signal propagation. The
magnitude of this effect has not been investigated. Also, the finite size of
the transmitting antenna also leads to a code edge that is not sharp \cite%
{KrallBahder2001}.

\section{Pseudorange Measurement}

\begin{figure}
\includegraphics{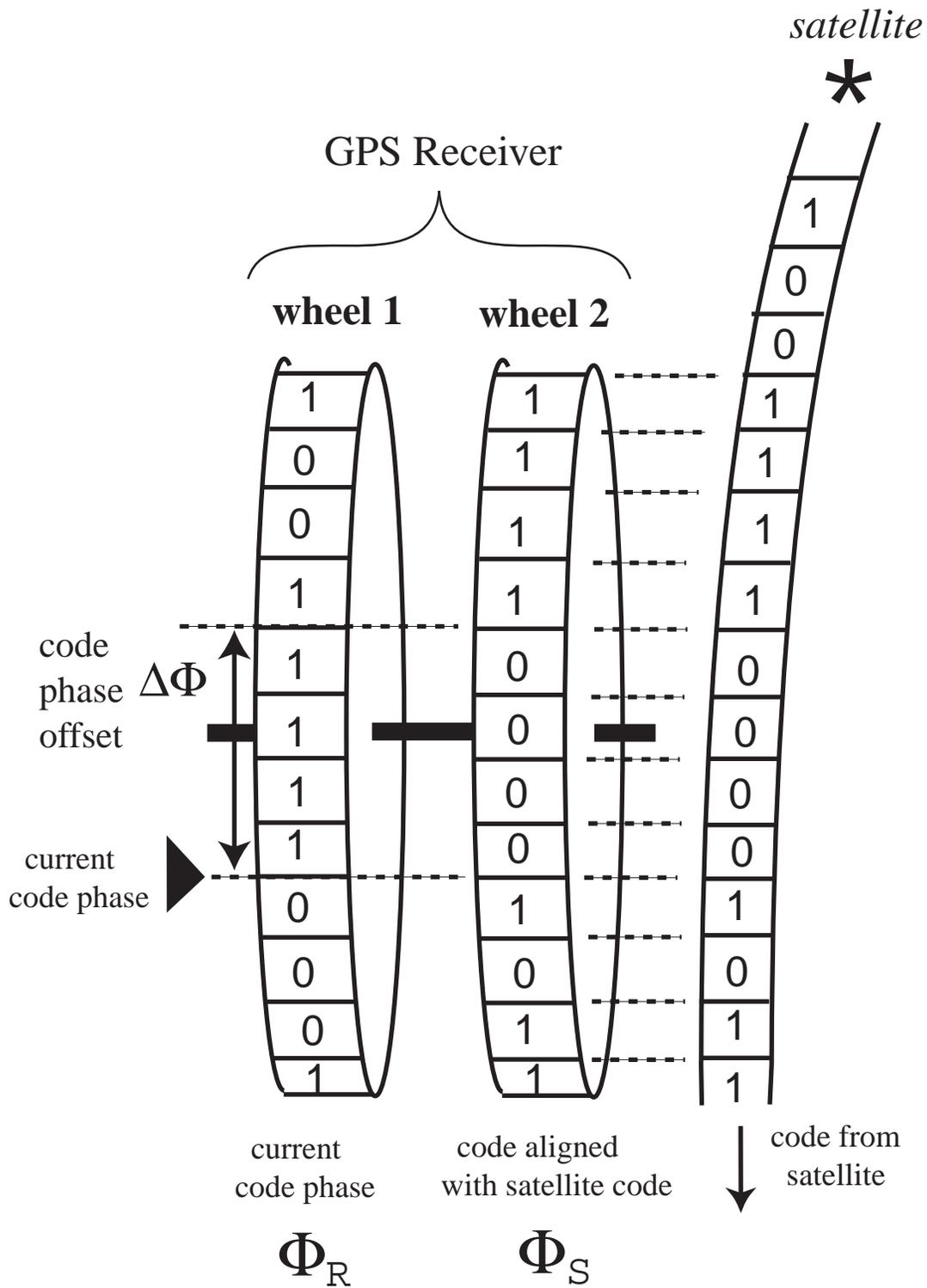}
\caption{\label{GPSreceiverCodeLock3} A mechanical analogue of a
GPS receiver tracks the code broadcast from a GPS satellite. The
receiver has two code wheels.  The phase angle of code wheel 1
represents the current time in the receiver.  The phase angle of
code wheel 2 is kept aligned with the in-coming code sequence from
the satellite.}
\end{figure}

A GPS receiver makes a special type of measurement, called a pseudorange
measurement, in which the Doppler and gravitational frequency shifts are
essentially removed (up to an additive constant) from the measurement. A GPS
receiver internally replicates (generates) the satellite P-code $s(t)$\ for
each satellite. The receiver determines which satellite signal it is
receiving by matching the internally replicated P-code with the code
received from the satellite. In what follows, we describe the pseudorange
measurement made by a GPS receiver by using a mechanical analogue. Our
mechanical analogue of a GPS receiver has two code wheels, numbered 1 and 2,
see Fig.~\ref{GPSreceiverCodeLock3}. During operation of the receiver, both
code wheels advance at the same angular velocity, $\Omega_{R}$. \ The
receiver will align the code incoming from the satellite with the code on
wheel 2. The code arriving is like a bicycle chain that is to be placed onto
the sprocket (code wheel 2). \ In order to align the codes, the receiver
must match the rate (frequency) at which bits are arriving as well as match
the duration of each bit. In general, the receiver has a non-zero velocity
in the ECI frame and it is at a different gravitational potential than the
transmitting satellite. At any instant in the receiver's comoving frame, the
frequency of code bits received from the satellite is $f_{R}=\omega_{R}/2\pi$%
, where $f_{R}$ is the number of received bits per second of the code $%
s(\tau^{\ast})$. To align the codes, this frequency $\omega_{R}$ must be
made equal to $\omega_{o},$ which is the frequency seen in the comoving
frame of the receiver, given in Eq.(\ref{observedDoppler}).

The temporal duration of the bits on code wheel 2 and the temporal duration
of incoming bits from the satellite are made equal by adjusting the radius $%
r_{w}$ of code wheel 2. \ Code wheel \ 1 is adjusted to the same radius as
code wheel 2. If the code wheel has a circumference equal to $2\pi r_{w}$,
the length per bit along the the circumference is $2\pi r_{w}/N$, where
there are $N$ bits in the code. In order to match incoming bits, the code
wheel must have a linear velocity $2\pi r_{w}f_{R}/N=r_{w}\Omega_{R}$. The
angular velocity of the code wheel is then $\Omega_{R}=\omega_{R}/N$. This
angular velocity of the code wheel, and consequently the frequency at which
the replicated code is generated, $\omega_{R}$, will be adjusted continually
and depends on the motion of the receiver and satellite, so that the
replicated code is constantly aligned with the code arriving from the
satellite:
\begin{equation}
\omega_{R}=\omega_{o}
\end{equation}
Therefore, a GPS receiver searches in time (aligns code phase on wheel 2
with incoming code) and in frequency (adjustment of $r_{w}$ to make temporal
length of bits on wheel 2 equal to satellite bits). This alignment of the
code is commonly referred to as \textquotedblleft code
tracking\textquotedblright\ or \textquotedblleft code lock\textquotedblright.

At any instant in time, code wheel 1 has a current value of phase, $\Phi_{R}
$, which is indicated by the pointer, see Fig. 4. The value of the
replicated phase $\Phi_{R}$ on code wheel 1 is a representation of the
receiver's current hardware time, $\tau_{R}^{\ast}$. The phase $\Phi_{R}$
and receiver clock hardware time $\tau_{R}^{\ast}$ are related by
\begin{equation}
\Phi_{R}=\frac{\omega_{s}}{N}\tau_{R}^{\ast}
\label{ReceiverPhaseHarwdareTime}
\end{equation}
where $\omega_{s}$\ is a system constant given by Eq.(\ref{FactoryOffset}).
The system constant $\omega_{s}$\ is a conversion factor from hardware time,
$\tau_{R}^{\ast}$, to phase $\Phi_{R}$. A key point is that during
\textquotedblleft code tracking\textquotedblright\ the phase $\Phi_{R}$ on
code wheel 2 is adjusted to keep pace with the incoming code from the
satellite, but this is done by advancing the time $\tau_{R}^{\ast}$, and not
by changing the frequency $\omega_{s}$, which is a fixed constant in the
system.

The replicated phase in the receiver can be related to the receiver's world
line by writing the hardware time $\tau_{R}^{\ast}$ in terms of the
coordinate time along the receiver's world line in the ECI frame, $x_{R}^{0}$%
,\
\begin{equation}
x_{R}^{0}=c(\tau_{R}^{\ast}+\Delta\tau_{R}^{\ast})
\label{receiver clock bias}
\end{equation}
\ where \ $\Delta\tau_{R}^{\ast}$\ is the time correction to the receiver's
hardware time, commonly called a clock bias. \ The receiver time, $\tau
_{R}^{\ast}=\eta T/N$, is periodic, $\tau_{R}^{\ast}+T=\tau_{R}^{\ast}$,
where $T=1$ week (the period of the P-code $s(\tau_{R}^{\ast})$), and $\eta$
is a real number $0\leq\eta\leq N$, which corresponds to the interpolated
value of the bit number $n$ in the P-code $s(\tau_{R}^{\ast})$, see Fig.\ref%
{GPSreceiverCodeLock3}. \ This periodicity in $\tau_{R}^{\ast}$\ does not
introduce any ambiguity because it is assumed that a GPS\ user is located
within one light week distance from the Earth.\ Using Eq.(\ref{receiver
clock bias}) the receiver phase, $\Phi_{R}$, can be related to the geometry
by%
\begin{equation}
\Phi_{R}=\frac{\omega_{s}}{cN}\left( x_{R}^{0}-c\Delta\tau_{R}^{\ast}\right)
\label{receiver phase}
\end{equation}

On code wheel 2, the receiver has the replicated code (for a given
satellite). This replicated code is rotated back and forth on code wheel 2
until it is aligned with the code phase $\Phi_{s}(t_{R},\mathbf{r}_{R})$\
arriving from the satellite, where $\mathbf{r}_{R}$ is the position of the
receiver. \ As described above, at all times the phase on code wheel 2 is
aligned with the incoming code phase, in both time (angle angle of the
wheel) and frequency or bit duration (radius $r_{w}$\ is adjusted).

The pseudorange measurement made by a GPS receiver at space-time point $%
(t_{R},\mathbf{r}_{R})$\ is the difference of the phases on code wheel 1 and
2~\cite{Ward-in-Kaplan}
\begin{equation}
\Delta\Phi(t_{R},\mathbf{r}_{R})\equiv\Phi_{R}-\Phi_{s}(t_{R},\mathbf{r}_{R})
\end{equation}
where $\Phi_{R}$ is the curent value of the replicated code phase in the
receiver (on code wheel 1) and $\Phi_{s}(t_{R},\mathbf{r}_{R})$ \ is the
value of the broadcast phase $\Phi(t,\mathbf{r})$ for satellite $s$ (given
in Eq.(\ref{Phase on hypersurface with clock offset})) evaluated at the
receiver at space-time measurement point $(t_{R},\mathbf{r}_{R})$. \ The
phase difference, $\Delta\Phi(t_{R},\mathbf{r}_{R})$, is a bonafide
measurement because it is a comparison (a difference) between the value of
the scalar field at the receiver, $\Phi_{s}(t_{R},\mathbf{r}_{R})$, and the
replicated code phase, $\Phi_{R}$. The measured difference, $%
\Delta\Phi(t_{R},\mathbf{r}_{R})$, is commonly expressed in units of length
and is called a pseudorange measurement from receiver to satellite $s$:
\begin{equation}
\rho_{s}=c\frac{N}{\omega_{s}}\left[ \Phi_{R}-\Phi_{s}(t_{R},\mathbf{r}_{R})%
\right]  \label{pseudorangeMeasurement}
\end{equation}
where $\rho_{s}$\ is the measured pseudorange in units of length.

The measured pseudorange,\ $\rho_{s}$, is a scalar under
generalized coordinate transformations (e.g., under Lorentz
transformations) because the phase function
$\Phi_{s}(t,\mathbf{r})$ is a scalar field and $\Phi_{R}$ is an
invariant that depends on the world line of the receiver. The
pseudorange depends on both the world line of the satellite and
the world line of the receiver, which are
invariant quantities. The world line of satellite $s$, $x_{s}^{i}(\tau^{%
\ast})$\thinspace,\ enters into the definition of the scalar field
$\Phi _{s}(t,\mathbf{r})$, see Eq.(\ref{Phase on hypersurface with
clock offset}). The world line of the receiver enters into the
value of the phase of the replicated code, $\Phi_{R}$, see
Eq.(\ref{receiver phase}).  The world line of the receiver also
enters into the evaluation of the scalar field $\Phi
_{s}(t,\mathbf{r})$  at the position of the receiver at event $R=(t_{R},%
\mathbf{r}_{R})$.  Therefore, the pseudorange measurement,
$\rho_{s}$, is a two-point scalar that depends on the emission
event $T$ and the reception event $R$. The pseudorange is a
particular case of quantities known as two-point tensors, which
depend on two space-time points, and have tensorial transformation
properties with respect to generalized coordinate transformations
(that can differ) at each point\cite{Synge1960}. Therefore, the
pseudorange $\rho_{s}$ is a two-point scalar under separate
generalized
coordinate transformations at $T=(t_{T},\mathbf{r}_{T})$ and at $R=(t_{R},%
\mathbf{r}_{R})$ and can be labelled as $\rho(t_{T},\mathbf{r}_{T},t_{R},%
\mathbf{r}_{R})$\cite{Synge1960}.\ \ \ The world function of space-time, $%
\Omega(x_{1}^{i},x_{2}^{j})$ given in Eq.(\ref{WorldFunction}), \
is a well-known two-point scalar, and not surprisingly, the
pseudorange and the world function have the same transformation
properties.

\ Two point-scalar fields enter into measurement situations
whenever there is a field generated at space-time point $T$, such
as an electromagnetic field, and a measurement or
\textit{projection} is done at the measurement event at space-time
point $R$. \ In this sense, there is some similarity between a
relativistic treatment of measurement processes and a quantum
mechanical treatment, which is mentioned in the introduction. \ In
the case of relativity, a measurement of a field is a projection
of the field onto an observer's tetrad \cite{Pirani57,Synge1960}.
\ In quantum mechanics, traditionally projection operators are
invoked in the definition of measurement
\cite{vonNeumann55,Omnes-book1994,Preskill-notes1998}.

\section{Time Transfer and Navigation in Space-Time}

The measured pseudorange, $\rho_{s}$ in Eq.(\ref{pseudorangeMeasurement}),
can be related to the geometric range between the emission event $T=(t_{T},%
\mathbf{r}_{T})$ and the reception event $R=(t_{R},\mathbf{r}_{R})$ by using
Eq.(\ref{receiver phase}) for $\Phi_{R}$ and evaluating the broadcast phase $%
\Phi(t,\mathbf{r})$ for satellite $s$ (given in Eq.(\ref{Phase on
hypersurface with clock offset})) at the reception event $R$, leading to

\begin{equation}
\rho_{s}=|\mathbf{r}_{R}\mathbf{-r}_{T}|+c\Delta\tau_{T}^{\ast}-c\Delta
\tau_{R}^{\ast}+\Delta(\mathbf{r}_{T},\mathbf{r}_{R})
\label{Measured Pseudorange}
\end{equation}
where we have set $\Delta\tau_{T}^{\ast}=\Delta\tau_{\eta}^{\ast}$. In Eq.(%
\ref{Measured Pseudorange}), $|\mathbf{r}_{R}\mathbf{-r}_{T}|$ is\ the
geometric range between events $T$ and $R$, $\Delta\tau_{R}^{\ast}$ is the
clock correction to the receiver clock at event $R$, and $c\Delta\tau
_{T}^{\ast}=x_{s}^{0}(T)-$\ $c\tau_{s}^{\ast}(T)$ is the satellite clock
correction at event $T$. The satellite clock correction, $%
\Delta\tau_{T}^{\ast}$, is broadcast by the satellite in the navigation
message contained in the function $D(\tau^{\ast})$, see Eq.(\ref%
{ElectricField}). \ Conventionally, the pseudorange is modelled\cite%
{ParkinsonGPSReview,Hofmann-Wellenhof93,Kaplan96} in flat space-time as $%
\rho_{s}=|\mathbf{r}_{R}\mathbf{-r}_{T}|+c\Delta\tau_{T}^{\ast
}-c\Delta\tau_{R}^{\ast}$. \ The term $\Delta(\mathbf{r}_{T},\mathbf{r}_{R})$%
\ is a small correction due to the presence of the Earth's gravitational
field that modifies space-time geometry near Earth. \ This correction
depends on the mass of the Earth, $M$, the angular velocity of Earth
rotation, $\Omega$, and the Earth's equatorial radius, $R$, and is given by%
\begin{equation}
\Delta(\mathbf{r}_{T},\mathbf{r}_{R})=\frac{2GM}{c^{2}\,\,}\,\left( \Lambda(%
\mathbf{r}_{T},\mathbf{r}_{R})-\frac{|\mathbf{r}_{R}\mathbf{-r}_{T}|}{R}%
\right) -\frac{\Omega^{2}R^{2}}{c^{2}}|\mathbf{r}_{R}\mathbf{-r}_{T}|
\label{pseudorange GR correction}
\end{equation}
where the purely geometric function $\Lambda(\mathbf{r}_{T},\mathbf{r}_{R})$%
\ is given in Eq.(\ref{Lambda12 Function}). \ The quantity $%
2GM/c^{2}\approx0.89$ cm is the gravitational radius of the
Earth,\ and
$\Omega^{2}R^{2}|\mathbf{r}_{R}\mathbf{-r}_{T}|/c^{2}\approx4.8
\times10^{-5}$ m, using $|\mathbf{r}_{R}\mathbf{-r}_{T}|\approx
a-R$, where $a$ is the semimajor axis of the GPS satellites. The
pseudorange in Eq.(\ref{Measured Pseudorange}) is based on the
metric in the ECI frame of reference, which is given in
Eq.(\ref{EarthMetricECI}).

If a user of the GPS knows their spatial coordinates in the ECI frame, $%
\mathbf{r}_{R}$, then a single pseudorange measurement $\rho_{s}$ to one
satellite is sufficient to determine the users time $t_{R}$ from Eq.(\ref%
{Measured Pseudorange}). \ In Eq.(\ref{Measured Pseudorange}) $\mathbf{r}%
_{T}=\mathbf{r}_{s}(t_{T})$, where $\mathbf{r}_{s}(t_{T})$ is the broadcast
satellite ephemeris evaluated at transmit time $t_{T}$ that satisfies%
\begin{equation}
\Omega(t_{T},\mathbf{r}_{s}(t_{T}),t_{R},\mathbf{r}_{R})=0
\label{transmit time from worldFunction}
\end{equation}
where $\Omega(t_{T},\mathbf{r}_{T},t_{R},\mathbf{r}_{R})$ is the world
function given in Eq.(\ref{WorldFunction}). \ Therefore,\ Eqs.(\ref{Measured
Pseudorange}) and (\ref{transmit time from worldFunction}) can be solved
numerically for the user's reception event time $t_{R}$, which is a
coordinate time in the ECI metric in Eq.(\ref{EarthMetricECI}).

The more common case is that a user of the GPS wants to obtain time but he
knows his spatial coordinates in the ECEF frame of reference, $\mathbf{y}%
=(y^{1},y^{2},y^{3})$. The user's ECI frame coordinates, $\mathbf{r}%
_{R}=(x^{1},x^{2},x^{3})$, are given by

\begin{equation}
\mathbf{r}_{R}=D(t_{T}-t_{0})\cdot\mathbf{y}  \label{ECI- ECEF rotation}
\end{equation}
where \ $D(t-t_{0})$\ is a time-dependent rotation matrix, which is equal to
the unit matrix at epoch time $t=t_{0}$, see Eq.(\ref%
{CoordinateTransformation1}). \ \ In this case, to obtain the user's time $%
t_{R}$ at the reception event, $\mathbf{r}_{R}$ in Eq.(\ref{Measured
Pseudorange}) and (\ref{transmit time from worldFunction}) must be
eliminated by use of Eq.(\ref{ECI- ECEF rotation}). Equations (\ref{Measured
Pseudorange}), and (\ref{transmit time from worldFunction}) can be solved
for $t_{R}$.

Navigation in space-time means the simultaneous determination of user
position and time, i.e., determination of the space-time coordinates of the
reception event $R=(t_{R},\mathbf{r}_{R})$. \ Navigation can be carried out
by simultaneously measuring four pseudoranges, $\rho_{s}$, to four different
satellites,\ $s=1,2,3,4$. \ It is clear that four simultaneous equations of
the form in Eq.(\ref{Measured Pseudorange}) can be solved for the four
coordinates $(t_{R},\mathbf{r}_{R})$ that specify the user reception event
in the ECI frame. Typically, a GPS user on Earth wants to know their
coordinates in the ECEF\ frame of reference. \ The user must determine the
coordinate time of emission for one satellite from Eq.(\ref{transmit time
from worldFunction}) and use it in the transformation Eq.(\ref{ECI- ECEF
rotation}) to determine user ECEF coordinates $\mathbf{y}%
=(y^{1},y^{2},y^{3}) $ \cite{AshbyWeiss1999}.

\bigskip

\section{Resolution of Two Receiver Experiment}

If two receivers are at the same event $M$ in space-time, and they
are tracking the same four satellites, do the receiver's measure
the same pseudorange?  Also, do they obtain the correct space-time
coordinates for the event $M$? The resolution of the two receiver
thought experiment, which was described in Section II, is now
clear. The value of the phase received from each satellite,
$\Phi_{s}(t_{R},\mathbf{r}_{R})$, is the same for each receiver,
independent of their velocities, because $\Phi_{s}(t,\mathbf{r})$
is a scalar field.  However, the value of the hardware time for
each receiver, and therefore the current replicated code phase,
$\Phi_{R}$, is different for each receiver. \ This difference is
due to the different clock bias $\Delta\tau_{R}^{\ast}$\ of the
two receivers. We can then say that each measured pseudorange is
an invariant under coordinate transformations, however, the
pseudoranges (to the same satellite) are different for each
receiver because they depend on receiver world lines. For
navigation purposes, even though the set of pseudoranges are
different for each of the two receivers, the
clock bias for each receiver is determined in the course of solving the four Eq. (\ref%
{Measured Pseudorange}).  Therefore, while each receiver obtains a \textit{%
different} set of four invariant pseudoranges to the same four satellites,
each receiver obtains the correct space-time coordinates of the reception
event $M=(t_{R},\mathbf{r}_{R})$.

\section{Summary}

We used a two-receiver thought experiment in Section II to
motivate the need to understand the transformation properties of
the measured pseudorange, the  quantity that is measured in the
GPS.

Starting from the weak field approximation for the metric of
space-time in the vicinity of the Earth, we used the standard
transformation to the rotating frame, and defined a new coordinate
time so that coordinate time on the geoid surface is equal to
proper time. Then we transform from the ECEF frame to the ECI
frame, to obtain the metric that describes the ECI frame of
reference near the Earth, given in Eq.(\ref{EarthMetricECI}).  In
this metric, due to a combination of gravitational and time
dilation effects, the GPS satellite clocks appear to run fast with
respect to coordinate time. We described the \textquotedblleft
factory offset" that is routinely applied to slow down the
satellite clocks to (approximately) keep coordinate time in the
ECI frame.  In Section IV-B, for the metric in the ECI frame given
in Eq.(\ref{EarthMetricECI}), we derived the apparent frequency of
the satellite signal seen by an observer at an arbitrary position
moving with an arbitrary velocity, due to Doppler and
gravitational potential differences, see
Eq.(\ref{observedDoppler}).

We described the nature of the digital signal (P-code) that is
broadcast by GPS satellites, and the space-time grid created by
GPS satellites due to the discontinuities in the broadcast
electromagnetic field. We computed the world function of
space-time, given in Eq.(\ref{WorldFunction}), for the ECI metric
in Eq.(\ref{EarthMetricECI}). Using this world function, we
defined a scalar phase field $\Phi_{s}(t,\mathbf{r})$, given in
Eq.(\ref{Phase on hypersurface with clock offset}), that satisfies
the eikonal equation. This phase field is seen in the same way by
all observers, independent of their state of motion. Using this
phase field, and a mechanical analoque for
a GPS\ receiver, we defined the measured pseudorange in Eq.(\ref%
{pseudorangeMeasurement}), and found that under generalized
coordinate transformations it transforms as a two-point scalar,
just like the world function. Within the geometrical optics
approximation, there are no velocity effects on pseudorange
measurements. Finally, we related the measured pseudorange to the
geometry of space-time in Eq.(\ref{Measured Pseudorange}). \ We
obtained a small
correction, given by $\Delta(\mathbf{r}_{T},\mathbf{r}_{R})$ in Eq.(\ref%
{pseudorange GR correction}), to the conventional model of
pseudorange. This correction is due to the curvature of space-time
in the vicinity of the Earth.

\begin{acknowledgments}
This work was supported by the Advanced Research and Development
Activity (ARDA).
\end{acknowledgments}

\appendix

\section{Conventions and Notation}

We use the convention that Roman indices, such as found on space-time
coordinates $x^{i}$, take the values $i=0,1,2,3$, and Greek indices take
values $\alpha=1,2,3$. Summation is implied over the range of an index when
the same index appears in a lower and upper position. If $x^{i}$ and $%
x^{i}+dx^{i}$ are two events along the world line of an ideal clock, then
the proper time interval between these events is $d\tau=ds/c$, where $ds$ is
given in terms of the space-time metric as $ds^{2}=-g_{ij}\,dx^{i}\,dx^{j} $%
. We choose $g_{ij}$ to have the signature $+$2, so when $g_{ij}$ is
diagonalized at any given space-time point, the elements can take the form
of the Minkowski metric given by $\eta_{00}=-1$, $\eta_{\alpha\beta}=%
\delta_{\alpha\beta}$.

\section{Doppler Effect in Earth's Gravitational Field}

\begin{figure}
\includegraphics{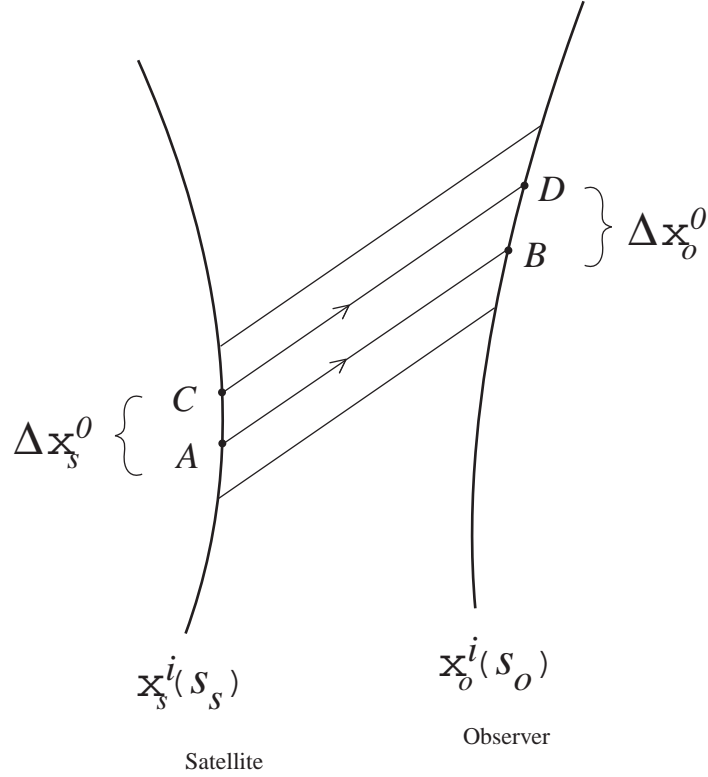}
\caption{\label{DopplerDerivationFig} The world line of satellite
and receiver are shown together with the 3-d hypersurfaces of
constant phase between point AB and CD.}
\end{figure}

A satellite at spatial position $\mathbf{r}_{s}$ and travelling at velocity $%
\mathbf{v}_{s}$ emits a signal \ in its comoving frame with frequency $%
\omega _{s}$. \ An observer at spatial position $\mathbf{r}_{o}$ and
travelling at velocity $\mathbf{v}_{o}$ measures the frequency of this
signal to be $\omega _{o}$ in his comoving frame. These frequencies are
related by Eq.(\ref{observedDoppler}). \ This relation can be derived by
evaluating \cite{Synge1960}
\begin{equation}
\frac{\omega _{o}}{\omega _{s}}=1-\frac{\Omega _{i_{s}}V^{i_{s}}+\Omega
_{i_{o}}V^{i_{o}}}{\Omega _{i_{s}}V^{i_{s}}}
\label{Doppler def - using world function}
\end{equation}%
where $\Omega _{i_{s}}$ and $\Omega _{i_{o}}$ are the covariant derivatives
of the world function with respect satellite and receiver coordinates, $%
x_{s}^{i}$\ and \ $x_{o}^{j}$, respectively, and\ $V^{i_{s}}$ and $V^{i_{o}}$
are the components of the 4-velocities of satellite and receiver at
space-time points $S=x_{s}^{j}=(t_{T},\mathbf{r}_{T})$ and $%
O=x_{o}^{i}=(t_{R},\mathbf{r}_{R})$, respectively. The form of Eq.(\ref%
{Doppler def - using world function}) together with the transformation
properties of the world function, shows that the quantity $\omega
_{o}/\omega _{s}$ is a two-point scalar under generalized coordinate
transformations.\ \ We have computed the world function in Eq. (\ref{Doppler
def - using world function}), $\Omega =\Omega (x_{s}^{i},x_{o}^{j})$, and it
is given in Eq.(\ref{WorldFunction}) for the metric given in Eq.(\ref%
{EarthMetricECI}). \ \ To obtain an explicit expression for $\omega
_{o}/\omega _{s}$ we use an alternative derivation \cite{Synge1960}. \
Consider the satellite and observer world lines, $x_{s}^{i}(s_{s})$ and $%
x_{o}^{i}(s_{o})$, which are parametrized by proper times $s_{s}/c$ and $%
s_{o}/c$, \ respectively, \ see Fig.\ref{DopplerDerivationFig}. \ The lines $%
AB$ and $CD$ are null geodesics that connect satellite and receiver world
lines, at successive cycles of the emitted signal.

The frequency ratio $\omega _{o}/\omega _{s}$\ is related to the
proper times

\begin{equation}
\frac{\omega _{o}}{\omega _{s}}=\frac{ds_{s}}{ds_{o}}
\end{equation}
where $ds_{s}$ is the proper time between $AC$ and $ds_{o}$\ is
the proper time between $BD$. Using the metric in
Eq.(\ref{EarthMetricECI}), the ratio
of frequencies can be written as%
\begin{equation}
\frac{\omega _{o}}{\omega _{s}}=\frac{ds_{s}}{ds_{o}}=\left[ \frac{%
g_{00}(S)+g_{\alpha \beta }(S)v_{s}^{\alpha }v_{s}^{\beta }\,}{%
g_{00}(O)+g_{\alpha \beta }(O)v_{o}^{\alpha }v_{o}^{\beta }}\right] ^{1/2}%
\frac{\Delta x_{s}^{0}}{\Delta x_{o}^{0}}  \label{freqRatio2}
\end{equation}%
where the emission and reception events are connected by a null geodesic, so
that $\Omega (S,O)=0$. \ In Eq. (\ref{freqRatio2}), $g_{\alpha \beta }(S)$
is the metric on the satellite world line at event $S$, $\ g_{\alpha \beta
}(O)$ is the metric on the observer world line at event $O$, and $%
v_{s}^{\alpha }$ and $v_{o}^{\alpha }$ are the velocity components
of satellite and receiver, for $\alpha =1,2,3$. For example, the
satellite velocity is $v_{s}^{\alpha }=\Delta x_{s}^{\alpha
}/\Delta x_{s}^{0}$, where $\Delta x_{s}^{0}$ is the coordinate
time between events $AC$ on the satellite world line. \ In the
limit $\Delta x_{s}^{0}\rightarrow 0$ (and
therefore $\Delta x_{o}^{0}\rightarrow 0$), using the world function in Eq.(%
\ref{WorldFunction}) to connect points $AC$ and $BD$, a lengthy calculation
of Eq.(\ref{freqRatio2}) using the metric in Eq.(\ref{EarthMetricECI}) leads
to Eq.(\ref{observedDoppler}). The term $\Delta x_{s}^{0}/\Delta x_{o}^{0}$
in Eq.(\ref{freqRatio2}) leads to linear velocity terms in Eq.(\ref%
{observedDoppler}).


\bigskip

\begin{thebibliography}{99}
\bibitem{Chuang2000} I. L. Chuang, \textquotedblleft Quantum Algorithm for
Distributed Clock Synchronization", Phys. Rev. Lett. \textbf{85}, 2006
(2000), and also in quant-ph/0004105.

\bibitem{Jozsa2000} R. Jozsa, D. S. Abrams, J. P. Dowling, and C. P.
Williams, \textquotedblleft Quantum Clock Synchronization Based on Shared
Prior Entanglement", Phys. Rev. Lett. \textbf{85}, 2006 (2000).

\bibitem{Preskill2000} J. Preskill, \textquotedblleft Quantum Clock
Synchronization and Quantum Error Correction", quant-ph/0010098.

\bibitem{Giovannetti2001} V. Giovannetti, S. Lloyd, and L. Maccone, Nature
\textbf{412}, 417-419 (2001);

\bibitem{ShihPRL2003} Y. Shih, Phys. Rev. Lett., in publication.

\bibitem{Burt2001} E. A. Burt, C. R. Ekstrom, T. B. Swanson,
\textquotedblleft Comment on \textquotedblleft Quantum Clock Synchronization
Based on Shared Prior Entanglement"", Phys. Rev. Lett. \textbf{87}, 129801
(2001); \textquotedblleft A Reply to \textquotedblleft Quantum Clock
Synchronization"", quant-ph/0007030.

\bibitem{Jozsa2000a} R. Jozsa, D. S. Abrams, J. P. Dowling, and C. P.
Williams, \textquotedblleft Jozsa et al. Reply", Phys. Rev. Lett. \textbf{87}%
, 129802 (2001).

\bibitem{Yurtsever2000} U. Yurtsever and J. P. Dowling, \textquotedblleft A
Lorentz-invariant Look at Quantum Clock Synchronization Protocols Based on
Distributed Entanglement", quant-ph/0010097.

\bibitem{ParkinsonGPSReview} B. W. Parkinson and J. J. Spilker, eds., Global
Positioning System: Theory and Applications, vol. I and II (P. Zarchan,
editor-in-chief), Progress in Astronautics and Aeronautics, vol. 163 and 164
(Amer. Inst. Aero. Astro., Washington, D.C., 1996).

\bibitem{Kaplan96} E. D. Kaplan, Understanding GPS: Principles and
Applications, Mobile Communications Series (Artech House, Boston, 1996).

\bibitem{Hofmann-Wellenhof93} B. Hofmann-Wellenhof, H. Lichtenegger, and J.
Collins, Global Positioning System Theory and Practice (Springer-Verlag, New
York, 1993).

\bibitem{GLONASS} The Russian satellite system known as GLONASS is similar
to GPS.

\bibitem{Rovelli2002} C. Rovelli, Phys.Rev. D65 (2002) 044017.

\bibitem{Blagojevic2002} M. Blagojevi, J. Garecki, F. W. Hehl, and Yu. N.
Obukhov, Phys. Rev. D 65, 044018 (2002).

\bibitem{KrallBahder2001} A. H. Krall and T. B. Bahder, J. Appl. Phys.
\textbf{90}, 6513 (2001).

\bibitem{RelativisticEffects} Actually there are three main relativistic
effects that contribute significantly, see section II.

\bibitem{LLClassicalFields} L. D. Landau and E. M. Lifshitz, \textit{%
Classical Theory of Fields}, Pergamon Press, New York, Fourth Revised
English Edition, (1975).

\bibitem{Caputo1967} M. Caputo, \textit{The Gravity Field of the Earth}
(Academic Press, N. Y., 1967).

\bibitem{AshbyInParkinsonGPSReview} N. Ashby and J. J. Spilker,
\textquotedblleft Introduction to Relativistic Effects on the Global
Positioning System", in \textit{Global Positioning System: Theory and
Applications}, B. W. Parkinson and J. J. Spilker, eds., vol. I and II (P.
Zarchan, editor-in-chief), Progress in Astronautics and Aeronautics, vol.
163 and 164 (Amer. Inst. Aero. Astro., Washington, D.C., 1996).

\bibitem{JGM-2constants} R. S. Nerem, F. J. Lerch, J. A. Marshall, E. C.
Pavlis, B. H. Putney, B. D. Tapley, R. J. Eanes, J. C. Ries, B. E. Schutz,
C. K. Shum, M. M. Watkins, S. M. Klosko, J. C. Chan, S. B. Luthcke, G. B.
Patel, N. K. Pavlis, R. G. Williamson, R. H. Rapp, R. Biancale, and F.
Nouel, \textquotedblleft Gravity model development for TOPEX/POSEIDON: joint
gravity models 1 and 2", J. Geophys. Res., 99(C12), 24421-24447, 1994b.

\bibitem{SpilkerInParkinsonGPSReview} See J. J. Spilker, \textquotedblleft
Chapter 3: GPS Signal Structure and Theoretical Performance", in \textit{%
Global Positioning System: Theory and Applications}, B. W. Parkinson and J.
J. Spilker, eds., vol. I and II (P. Zarchan, editor-in-chief), Progress in
Astronautics and Aeronautics, vol. 163 and 164 (Amer. Inst. Aero. Astro.,
Washington, D.C., 1996).

\bibitem{ICD-GPS-200} Anonymous, \textquotedblleft Navstar GPS Space
Segment/Navigation User Interfaces", ICD-GPS-200, Revision C, Initial
Release, ARINC Research Corporation, 10 October 1993.

\bibitem{Synge1960} J. L. Synge, \textit{Relativity: The General Theory},
North-Holland Publishing Co., New York, (1960).

\bibitem{Kheyfets1991} A. Kheyfets, \textquotedblleft Spacetime geodesy",
Weapons Laboratory, Air Force Systems Command, Kirtland Air Force Base, New
Mexico, Technical Report No. WL-TN-90-13, July 1991.

\bibitem{eSinE} This statement neglects the correction needed due to the
eccentricity of the satellite orbit, the so-called \textquotedblleft$e\sin E
$\ effect\textquotedblright.

\bibitem{Fock1964} V. Fock, \textquotedblleft The Theory of Space, Time and
Gravitation", Pergamon Press, Oxford, England, 1964.

\bibitem{DeFeliceClarke1990} F. De Felice and C. J. S. Clarke, \textit{%
Relativity on Curved manifolds}, Cambridge Monographs on Mathematical
Physics, Cambridge University Press, New York, 1990.

\bibitem{Bahder2001} T. B. Bahder, ``Navigation in curved space-time", Am.
J. Phys. \textbf{69}, 315 (2001).

\bibitem{projection on tetrad} The evaluation of a scalar field $\Phi _{s}(t,%
\mathbf{r})$\ at a point of observation (at the receiver), $(t_{R},\mathbf{r}%
_{R})$, is the same measurement operation as the projection of a tensor
field $F_{ij}$ onto an observer tetrad $\lambda_{(a)}^{i}$, where $i$ labels
the coordinate vector components and $a=1,2,3,4$, labels the tetrad basis
vectors. For the tensor field, the projection is done by the formula, $%
F_{(ab)}=F_{ij}$ $\lambda_{(a)}^{i}\lambda_{(a)}^{j}$, where $F_{(ab)}$\ are
invariants under coordinate transformations, but they depend on the observer
world line through the definition of the tetrad, see Ref. \cite{Synge1960}.

\bibitem{Ward-in-Kaplan} P. Ward, \textquotedblleft Satellite Signal
Acquisition and Tracking", Ch. 5 in E. D. Kaplan, \textit{Understanding GPS:
Principles and Applications}, Mobile Communications Series (Artech House,
Boston, 1996).

\bibitem{Pirani57} F. A. E. Pirani, Bulletin De L'Academie Polonaise Des
Sciences Cl. III, 1957, Vol. V, No. 2, p. 143--146 (1957).

\bibitem{vonNeumann55} J. von Neumann, \textit{Mathematical Foundations of
Quantum Mechanics}, Princeton University Press, Princeton, 1955.

\bibitem{Omnes-book1994} R. Omnes, \textit{The Interpretation of Quantum
Mechanics}, Princeton University Press, Princeton, 1994.

\bibitem{Preskill-notes1998} See also the discussion in J. Preskill, \textit{%
Lecture Notes for Physics 229: Quantum Information and Computation},
California Institute of Technology, 1998.

\bibitem{AshbyWeiss1999} N. Ashby and M. Weiss, \textquotedblleft Global
Positioning System Receivers and Relativity\textquotedblright, NIST\
Technical Note 1385, U.S. Dept. of Commerce, U.S. Government Printing
Office, Washington, 1999.
\end{thebibliography}
\end{document}